\documentclass[amsmath,twocolumn,amssymb,prb,aps,superscriptaddress,letterpaper,showpacs]{revtex4}

\usepackage{graphicx}
\usepackage{dcolumn}
\usepackage{bm}
\usepackage{amsmath}

\begin{document}

\title{Hyperfine interaction and electron-spin decoherence
in graphene and carbon nanotube quantum dots}  %

\author{Jan Fischer}
\affiliation{Department of Physics, University of Basel, Klingelbergstrasse 82,
4056 Basel, Switzerland}

\author{Bj{\"o}rn Trauzettel}
\affiliation{Institute of Theoretical Physics and Astrophysics,
University of W{\"u}rzburg, D-97074 W{\"u}rzburg, Germany}

\author{Daniel Loss}
\affiliation{Department of Physics, University of Basel, Klingelbergstrasse 82,
4056 Basel, Switzerland}

\date{\today}

\begin{abstract}
We analytically calculate the nuclear-spin interactions of a single electron 
confined to a carbon nanotube or graphene quantum dot. While the conduction-band 
states in graphene are $p$-type, the accordant states in a carbon nanotube are
$sp$-hybridized due to curvature. This leads to an interesting interplay between
isotropic and anisotropic hyperfine interactions. By using only analytical
methods, we are able to show how the interaction strength depends on important
physical parameters, such as curvature and isotope abundances. 
We show that for the investigated carbon structures, the $^{13}\mathrm{C}$ 
hyperfine coupling strength
is less than $1 \, \mu e\mathrm{V}$, and that the associated electron-spin
decoherence time can be expected to be several tens of microseconds or longer, depending
on the abundance of spin-carrying $^{13}\mathrm{C}$ nuclei. 
Furthermore, we find that the hyperfine-induced Knight shift is highly
anisotropic, both in graphene and in nanotubes of arbitrary chirality.
\end{abstract}


\maketitle

\section{Introduction}\label{sec:intro}

The spin of an electron confined to a semiconductor quantum dot is a prime candidate for
quantum information processing devices, with potential applications in
spintronics \cite{AwschalomBook, Zutic2004, Awschalom2007} and quantum 
computation. \cite{Loss1998, Cerletti2005, Hanson2007}
Schemes for quantum computation rely on a sufficiently long lifetime
of initialized spin states. One major problem is that the electron is not
isolated from its environment but interacts with the nuclei in the semiconductor
it has been confined to. This interaction leads to relaxation of excited
spin states, as well as to the decay of spin-state superpositions (decoherence).
For most quantum dots at low temperatures, the main source of spin
decoherence is the electron's interaction with a fluctuating magnetic field 
created by the randomly precessing nuclear spins. 
\cite{Burkard1999, Merkulov2002, Khaetskii2002, Coish2004, Johnson2005, Koppens2005,
Cywinski2009}
For III-V semiconductor quantum dots,
with their high abundance of spin-carrying nuclear isotopes,
the associated decoherence times are quite short, usually of order nanoseconds,
\cite{Merkulov2002, Khaetskii2002, Coish2004, Petta2005}
if no manipulations on the nuclear system are performed.

One way of overcoming the problem of short decoherence times
is to build quantum dots from semiconductors with lower abundances of spin-carrying 
isotopes, potentially resulting in weaker nuclear-spin interactions.
Carbon structures naturally consist of 99\% $^{12}\mathrm{C}$ with nuclear spin $0$ and only
of 1\% $^{13}\mathrm{C}$ with nuclear spin $\frac{1}{2}$, and are therefore 
promising materials for building quantum dots featuring long spin decoherence times.
This extraordinary property of carbon materials, as well as their
comparatively weak spin-orbit interactions, has led to
proposals of fabricating quantum dots in graphene ribbons with armchair boundaries 
\cite{Trauzettel2007} and in carbon nanotubes (CNTs). \cite{Bulaev2008}

\begin{figure}[t]
  \centering
  \includegraphics[width=0.95\columnwidth]{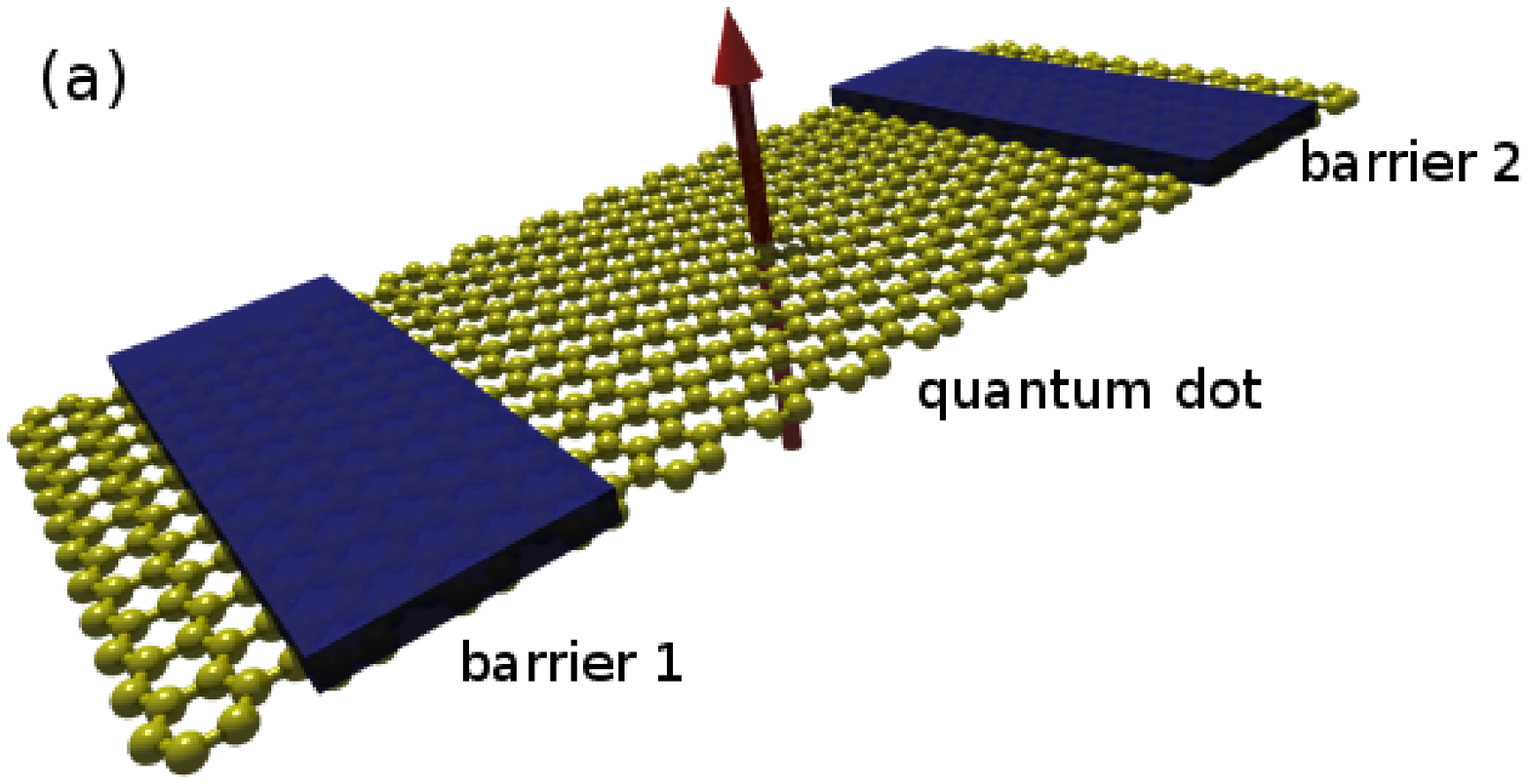}\\
  \includegraphics[width=0.95\columnwidth]{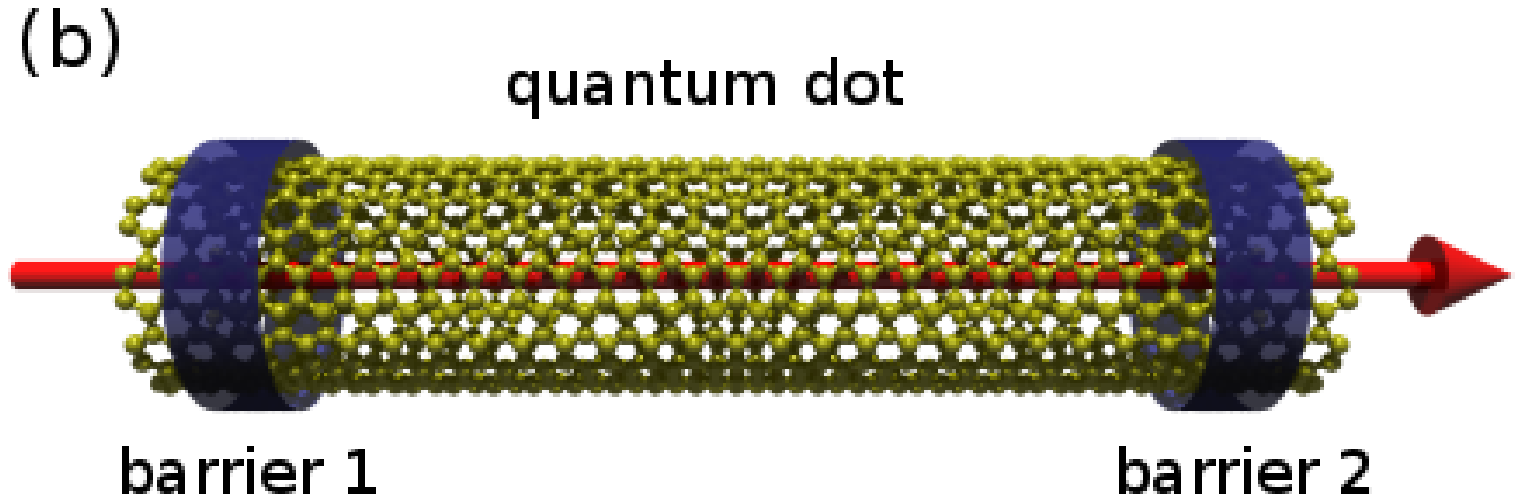}
  \caption{Sketch of the systems under consideration throughout this work: (a)
  a single electron confined to a quantum dot defined by two barriers on a 
  graphene ribbon; (b) a single-electron quantum dot defined in a carbon nanotube.
  Externally applied magnetic fields are indicated by the red arrows.
  }
  \label{fig:nanotube}
\end{figure}

Although it was not yet possible to experimentally realize a single-electron quantum 
dot in graphene, Coulomb-blockade measurements in gated graphene quantum dots
have already been carried out successfully. \cite{Schnez2008, Ponomarenko2008}
Experiments on single electrons in CNTs have advanced even more. 
\cite{Tans1997, Bockrath1997, Kong2000, Minot2004, Jarillo2004, Mason2004, Biercuk2005, 
Cao2005, Sapmaz2006, Onac2006, Graeber2006, Jorgensen2006, Meyer2007, Kuemmeth2008, Steele2009}
Very recently, first measurements on nuclear-spin interactions and electron-spin dynamics in  
CNTs have been performed, \cite{Churchill2009a, Churchill2009b} reporting an unexpectedly
strong hyperfine interaction of order $A \simeq 100 \, \mu e \mathrm{V}$
in $^{13}\mathrm{C}$ enriched CNTs (we will comment on the discrepancy between
this result and our calculations in Sec. \ref{sec:comparison}).
Furthermore, theoretical and experimental NMR studies on fullerenes 
\cite{Pennington1996} have been
carried out as well as \textit{ab initio} calculations on hyperfine interaction
in small graphene flakes. \cite{Yazyev2008} An analytical investigation of nuclear-spin 
interactions of electrons confined to graphene and CNTs, however, has so far been missing.

In our work, we analytically calculate the interaction of a single electron confined
to a graphene ribbon or a CNT (see Fig. \ref{fig:nanotube})
with the spins of the surrounding $^{13}\mathrm{C}$
nuclei. For CNTs we find an interesting interplay between isotropic and anisotropic 
hyperfine interactions which depends on the CNT geometry and which leads to 
a highly anisotropic Knight shift and an unusual alignment of the nuclear spins 
around a CNT circumference (in the ground state).
Furthermore, we calculate the decoherence dynamics 
of the electron spin and find that, even without manipulating the nuclear spins, 
the hyperfine-associated decoherence times can be on the order of
tens of microseconds or longer, depending on the relative abundance 
of $^{13}\mathrm{C}$ nuclei. These timescales are much longer than the ones typically
found for III-V semiconductor quantum dots, making carbon-based quantum
dots promising spin-qubit candidates.

Our paper is organized as follows: In Sec. \ref{sec:bonds} we determine the
wavefunction of a conduction-band electron in a CNT. 
In Sec. \ref{sec:interactions} we derive an effective Hamiltonian for the
nuclear-spin interactions and find a highly anisotropic coupling of the electron
to the $^{13}\mathrm{C}$ nuclear spins, which results in an anisotropic Knight shift
and an unusual alignment of the nuclear
spins around the nanotube, as shown in Sec. \ref{sec:knightshift}. 
In Sec. \ref{sec:dynamics}, we determine the electron-spin dynamics and the associated
spin decoherence times. We give a detailed comparison of our results with previous 
studies on comparable systems in Sec. \ref{sec:comparison}, and conclude in
Sec. \ref{sec:conclusions}.

\section{Bonds and bands}\label{sec:bonds}

The carbon atom features six electrons: two core electrons and four valence
electrons. In a solid, the four valence electrons (which are in $2s$ and $2p$ 
states) form the bonds with the
nearest-neighbor atoms. In two-dimensional graphite, or graphene, each atom
has three nearest neighbors, and three of the four valence electrons form
$sp^2$-hybridized bonds (called $\sigma$-bonds) with those neighbors,
while the fourth electron is in a so-called $\pi$-state perpendicular to 
the $\sigma$-bonds. \cite{SaitoBook} The $\pi$-electrons in graphene determine the
band structure near the Fermi energy, while the $\sigma$-electrons form more
remote bands. A conduction-band electron is therefore in a $\pi$-state.

\begin{figure}[t]
  \centering
  \includegraphics[width=0.95\columnwidth]{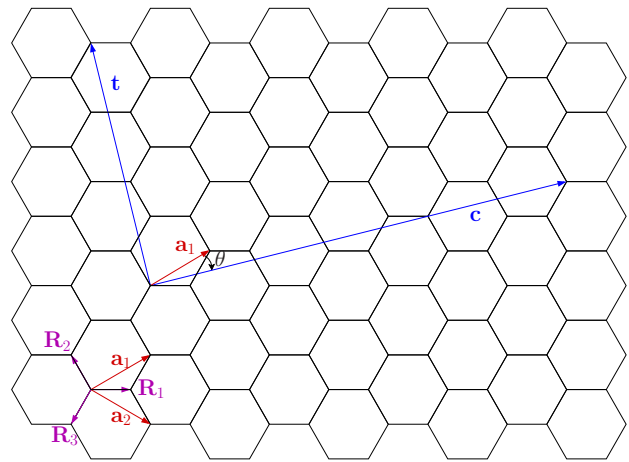}
  \caption{Definitions used throughout this work: $\mathbf{a}_1$ and $\mathbf{a}_2$
    denote the basis vectors of the honeycomb lattice, $\mathbf{R}_j$ are the 
    relative nearest-neighbor positions, $\mathbf{c}$ and $\mathbf{t}$ are the
    circumferential and transverse vectors, respectively, and $\theta = \theta_{nm}$ is the
    chiral angle between $\mathbf{c}$ and $\mathbf{a}_1$.}
  \label{fig:nt_defs}
\end{figure}

A CNT can be regarded as a graphene sheet that had been
rolled up along some direction $\mathbf{c}$, defining a symmetry axis
$\mathbf{t}$ of the CNT.
The circumferential and translational vectors are defined in terms of the basis
vectors $\mathbf{a}_1$ and $\mathbf{a}_2$ as (see Fig. \ref{fig:nt_defs})
\begin{equation}
  \mathbf{c} = n \mathbf{a}_1 + m \mathbf{a}_2, \quad
  \mathbf{t} = t_1 \mathbf{a}_1 + t_2 \mathbf{a}_2.
\end{equation}
Here, $n,m \in \mathbb{N}_0$ are the chiral indices, and $t_1 = (2m+n)/d_R$,
$t_2 = -(2n+m)/d_R$ with $d_R$ being the greatest common divisor of
$2n+m$ and $2m+n$. We denote the chiral angle between $\mathbf{c}$ and
$\mathbf{a}_1$ by $\theta=\theta_{nm}$.

The curvature of the CNT causes a geometrical tilting of the $\sigma$-bonds
which results in an $sp$-hybridization of the conduction-band states.
In Ref. \onlinecite{Kleiner2001}, this effect has been studied perturbatively 
in lowest order in the small parameter $2 \pi/L$, where 
$L=\sqrt{n^2+nm+m^2}$ is the circumference of the CNT in units of the lattice 
constant $a \simeq 2.5 \, \mathrm{\AA}$.
\footnote{For typical CNTs we have $2 \pi/L < 1$. However, the parameter can approach
unity for ultra-small nanotubes. At this point, our theory breaks down.}

The wavefunction of a conduction-band electron in a periodic crystal
is given by Bloch's theorem:
$\Psi_{\mathbf{k} \sigma} = \frac{1}{\sqrt{N_A}} e^{i \mathbf{k} \cdot \mathbf{r}}
u_{\mathbf{k} \sigma}(\mathbf{r})$, where $N_A$ is the number of atomic sites in 
the crystal and the Bloch amplitude $u_{\mathbf{k} \sigma}(\mathbf{r})$ has the periodicity 
of the lattice. Following Ref. \onlinecite{Fischer2008},
we approximate the Bloch amplitude $u_{\mathbf{k} \sigma}(\mathbf{r})$ at the
K and K' points by a linear combination of hydrogenic orbitals,
$u_{\sigma}(\mathbf{r}) = \sum_\mathbf{R} \pi(\mathbf{r}-\mathbf{R})$, where \cite{Kleiner2001}
\begin{align}
  \label{pi-orbital}
  \pi(\mathbf{r}) = &N_{nm} \Bigl\{ \psi_{2p_\perp}(\mathbf{r}) + 
    \frac{\pi}{2 \sqrt{3} L} \Bigl( \psi_{2s}(\mathbf{r}) \nonumber\\ 
    &+ \sin (3 \theta'_{nm}) \psi_{2p_t}(\mathbf{r}) + \cos (3 \theta'_{nm}) 
    \psi_{2p_c}(\mathbf{r}) \Bigr) \Bigr\},
\end{align}
and the sum runs over all lattice sites in the CNT. In the above,
$\psi_{2s}$ represents a hydrogenic  $2s$ orbital, $\psi_{2p_t}$, $\psi_{2p_c}$ 
and $\psi_{2p_\perp}$ are, respectively, the $2p$ orbitals along the transverse,
circumferential, and radial direction of the CNT, and $N_{nm}$ normalizes the Bloch 
amplitude to two atoms per unit cell.
$\theta'_{nm}$ is the angle between $\mathbf{c}$ and $\mathbf{R}_1$, and we can write
$\sin (3 \theta'_{nm}) = (n-m) (2n^2+5nm+2m^2)/2L^3$ and $\cos (3 \theta'_{nm}) 
= 3 \sqrt{3} n m (n+m)/2 L^3$ in terms of the chiral indices. \cite{Kleiner2001}
The hydrogenic orbitals consist of a radial and an angular part, e.g.,
$\psi_{2s}(\mathbf{r})=R_{20}(r) Y_0^0(\vartheta, \varphi)$. We will choose 
a local coordinate system at each lattice site, such that 
$\psi_{2p_t}$, $\psi_{2p_c}$, and $\psi_{2p_\perp}$ correspond to
hydrogenic $2p$ orbitals in $z$-, $y$-, and $x$-direction, respectively (see below).
The radial components of the hydrogenic orbitals depend on an effective 
screened nuclear charge $Z_{\mathrm{eff}}$ `seen' by the electron. \cite{Clementi1963}

Our choice of the Bloch amplitude implicitly assumes that the electron is tightly
bound to the nuclei, i.e., that the radial component of $\pi(\mathbf{r})$ drops off
fast on the scale of the nearest-neighbor distance. We have estimated that
$|\pi(\mathbf{r}+\mathbf{R}_{\mathrm{n.n.}})|^2/|\pi(\mathbf{r})|^2 \simeq 10^{-3}$
for any nearest-neighbor lattice vector $\mathbf{R}_{\mathrm{n.n.}}$, justifying
our assumption.

In a quantum dot, the electron is delocalized over many lattice sites and its
Bloch amplitude is modulated by an envelope function $\Phi_{\sigma}$ defined by the confinement
potential (see Ref. \onlinecite{Bulaev2008} for the envelope function of a quantum dot
defined by a rectangular confinement potential in a semiconducting CNT).
Including the spin states $|\sigma \rangle = |\uparrow, \downarrow \rangle$, 
we write the electron states as
\begin{equation}
  \label{spin-states}
  |\Psi_\sigma \rangle = |\Phi_{\sigma}; u_{\sigma} \rangle |\sigma \rangle,
\end{equation}
where, in the envelope-function approximation, $\langle \mathbf{r} | 
\Phi_{\sigma}; u_{\sigma} \rangle = \Phi_{\sigma}(\mathbf{r}) u_{\sigma}(\mathbf{r})$.
Since we did not include spin-orbit interactions in our model, the orbital
wavefunction is independent of the spin state, and we may omit the subscript
$\sigma$: $\Phi_\sigma(\mathbf{r}) = \Phi(\mathbf{r})$ and $u_\sigma(\mathbf{r}) 
= u(\mathbf{r})$.

\section{Nuclear-spin interactions}\label{sec:interactions}

\subsection{Carbon nanotubes}\label{sec:int_nanotube}

There are three terms that couple the spin of the confined electron to the nuclear
spins in the CNT: the Fermi contact interaction, the anisotropic hyperfine
interaction and the coupling of electron orbital angular momentum to the nuclear spins.
These interactions are represented by the Hamiltonians \cite{stoneham}
\begin{align}
  \label{ham:contact}
  h_1^k &= \frac{\mu_0}{4 \pi} \> \frac{8 \pi}{3} \> \gamma_S \gamma_{j_k} \>
  \delta(\mathbf{r}_{k}) \> \mathbf{S} \cdot \mathbf{I}_k,\\
  \label{ham:anisotropic}
  h_2^k &= \frac{\mu_0}{4 \pi} \> \gamma_S \gamma_{j_k} \> \frac{3 (\mathbf{n}_k \cdot 
  \mathbf{S}) (\mathbf{n}_k \cdot \mathbf{I}_k) - \mathbf{S} \cdot \mathbf{I}_k}
  {r_{k}^3 (1+d/r_{k})},\\
  \label{ham:angular}
  h_3^k &= \frac{\mu_0}{4 \pi} \> \gamma_S \gamma_{j_k} \> \frac{\mathbf{L}_k \cdot 
  \mathbf{I}_k} {r_{k}^3 (1+d/r_{k})},
\end{align}
respectively, where $\gamma_S=2\mu_B$, $\gamma_{j_k}=g_{j_k} \mu_N$, $\mu_B$ is the Bohr magneton,
$g_{j_k}$ is the nuclear g-factor of isotopic species $j_k$,
$\mu_N$ is the nuclear magneton, $\mu_0$ is the vacuum permeability,
$\mathbf{r}_{k} = \mathbf{r} - \mathbf{R}_k$ 
is the electron-spin position operator relative to the nucleus, 
$d \simeq Z \times 1.5 \times 10^{-15} \, \mathrm{m}$ is a length of nuclear dimensions, 
$Z$ is the charge of the nucleus, and $\mathbf{n}_k = \mathbf{r}_{k}/ r_{k}$. 
$\mathbf{S}$ and $\mathbf{L}_k = \mathbf{r}_k \times \mathbf{p}$ denote the spin and 
orbital angular-momentum operators (with respect to the $k^{\mathrm{th}}$ nucleus)
of the electron, respectively.
The cutoff $1+d/r_k$ comes from the Dirac equation (see, e.g., Ref. \onlinecite{stoneham})
and avoids unphysical divergences from expectation values of the Hamiltonians $h_2^k$ and $h_3^k$.
In the problem considered here, this cutoff may be omitted for the following reasons:
(i) The expectation values of $h_2^k$ and $h_3^k$ with respect to an $s$-state vanish 
identically due to the spherical symmetry of the wavefunction and the vanishing orbital
angular momentum, respectively. The expectation values with respect to a $p$-state are non-zero,
but the $p$-wavefunction goes to zero sufficiently fast at the position of each nucleus, 
thus avoiding a divergence.
(ii) As mentioned above, the electron wavefunction does not extend significantly to the
nearest-neighbor lattice sites. Hence, within our tight-binding approximation, the orbital 
$\pi(\mathbf{r})$ centered around some nucleus cannot cause a divergence at the position of 
a nearest neighbor.

We note that the orbital angular momentum $\mathbf{L}_k$ which appears
in Eq. \eqref{ham:angular} is associated with the electronic motion around the
$k^{\mathrm{th}}$ nucleus and is described by the Bloch part of the
electron wavefunction. On the other hand, it has been shown in Ref.
\onlinecite{Latil2001}, that interatomic currents can occur along the nanotube
circumference to which another orbital angular momentum $\mathcal{L}$ may be
associated which is described by the envelope part of the electron wavefunction.
However, the hyperfine coupling strength is defined via the Bloch part 
of the electron wavefunction
and therefore a consideration of the coupling between the valley degrees of freedom
via envelope-function-associated angular momenta is beyond the scope of the
present work.

We will consider CNTs and graphene with different abundances of the nuclear isotopes
$^{12}\mathrm{C}$ and $^{13}\mathrm{C}$. While $^{12}\mathrm{C}$ does not carry a nuclear
spin, the nuclear gyromagnetic ratio of $^{13}\mathrm{C}$ is non-vanishing and given
by $\gamma_{^{13}\mathrm{C}} = 7.1 \times 10^{-27} \> \mathrm{J}/\mathrm{T}$.

The Fermi contact interaction \eqref{ham:contact} yields a
finite contribution for $s$-states, but vanishes for $p$-states. 
The anisotropic hyperfine interaction \eqref{ham:anisotropic} and
the coupling of orbital angular momentum \eqref{ham:angular} vanish for $s$-states
because of their spherical symmetry and zero orbital angular momentum,
but yield a finite contribution for $p$-states.
Therefore, when considering CNTs, all three interactions \eqref{ham:contact} -
\eqref{ham:angular} have to be taken into account because of the $sp$-hybridized electron states
\eqref{pi-orbital}, while for graphene, only the interactions \eqref{ham:anisotropic} and
\eqref{ham:angular} are relevant, due to the purely $p$-type wavefunction (corresponding
to the limit $n,m \rightarrow \infty$ in Eq. \eqref{pi-orbital}).

We first calculate matrix elements of the interactions \eqref{ham:contact} 
- \eqref{ham:angular} with respect to the electron wavefunction 
\eqref{spin-states}, which will lead to effective spin
Hamiltonians and to the associated coupling strengths in the CNT case. 
Throughout this section, we will consider a CNT that consists only of spin-carrying 
$^{13}\mathrm{C}$ isotopes. The possibility of different nuclear isotope abundances will 
then be taken into account in Sec. \ref{sec:dynamics}.
From the CNT results, it will be possible to perform the `graphene limit', which we postpone
to Sec. \ref{sec:int_graphene}.

We start with the Fermi contact interaction and calculate
\begin{equation}
  \langle \Psi_{\sigma} | h_1^k | \Psi_{\sigma'} \rangle = \frac{2 \mu_0 \gamma_S \gamma_{^{13}\mathrm{C}}}{3} 
  \sum_k |u(\mathbf{r}_k)|^2 |\Phi(\mathbf{r}_k)|^2 \langle 
  \sigma | \mathbf{S} \cdot \mathbf{I}_k | \sigma' \rangle,
\end{equation}
assuming that the electron-spin density does not depend on the lattice site, which
is justified if, e.~g., the envelope function $\Phi(\mathbf{r})$ describes the ground
state of the quantum dot. The effects of a site dependence of the electron-spin
density have been recently considered in Ref. \onlinecite{Palyi2009}.
Evaluating the spin matrix elements leads to the following effective spin Hamiltonian:
\begin{equation}
  H_1 = \sum_k A_k^{(1)} \mathbf{S} \cdot \mathbf{I}_k,
\end{equation}
with coupling constants $A_k^{(1)} = A_1 v_0 |\Phi(\mathbf{r}_k)|^2$ 
(where $v_0$ is the volume of a primitive unit cell) and the associated
coupling strength
\begin{equation}
  \label{isotropic-coupling}
  A_1 = \frac{\mu_0 \gamma_S \gamma_{^{13}\mathrm{C}} Z_{\mathrm{eff}}^3}{3 \pi a_0^3} \> N_{nm}^2 \beta_{nm}^2,
\end{equation}
where we have introduced $\beta_{nm} = \pi/2 \sqrt{3} L$, and $a_0$ is the Bohr radius.
The normalization factor $N_{nm}$
can be determined by normalizing Eq. \eqref{pi-orbital} to two atoms per unit cell:
\begin{equation}
  N_{nm} = 2 \sqrt{\frac{L^2}{\pi^2+4L^2}}.
\end{equation}
We have evaluated Eq. \eqref{isotropic-coupling} for CNTs of different chiralities
in Table \ref{table:hyperfine}. For typical CNTs,
the coupling strength $A_1$ is about three orders of magnitude smaller than that for an electron
in a GaAs quantum dot ($A_1^{\mathrm{GaAs}} \simeq 90 \mu e \mathrm{V}$, see Ref. 
\onlinecite{Paget1977}), for two reasons: (i) The hybridization
prefactor $N_{nm} \beta_{nm}$ is on the order of $0.05$ and enters quadratically into
$A_1$. (ii) The effective nuclear charge (which
enters in third power into $A_1$) is $Z_{\mathrm{eff}}^{\mathrm{C}} \simeq 3.2$
for carbon, but $Z_{\mathrm{eff}}^{\mathrm{Ga}} \simeq 7.1$ for gallium and 
$Z_{\mathrm{eff}}^{\mathrm{As}} \simeq 8.9$ for arsenic. \cite{Clementi1963}

The sign of the isotropic interaction is positive for all nanotube diameters, 
in contrast to results reported previously, \cite{Semenov2007} 
where spin relaxation of conduction electrons has been calculated numerically and
a sign change of the hyperfine coupling constant $A_1$ has been found for small
nanotube diameters. This is due to the fact that in Ref. \onlinecite{Semenov2007} 
the coupling of the $1s$ core electrons to the nuclear spins
is included, which, however, is irrelevant for the hyperfine interaction of a 
conduction electron and, hence, for the electron spin dephasing considered in this article.

Now we look at the anisotropic hyperfine interaction. Due to symmetry, the $s$-part
of the hybridized wavefunction does not contribute. Taking matrix elements
$\langle \Psi_{\sigma} | h_2^k | \Psi_{\sigma'} \rangle$ just like above, we arrive
at an effective Hamiltonian
\begin{align}
	\label{H2}
    H_2 = \sum_k \Bigl( A_k^{(2,x)} S^x I_k^x + A_k^{(2,y)} S^y I_k^y + A_k^{(2,z)} S^z I_k^z \Bigr),
\end{align}
with coupling constants  $A_k^{(2,j)} = A_2^j v_0 |\Phi(\mathbf{r}_k)|^2$ and the coupling
strengths
\begin{equation}
  \label{anisotropic-coupling}
  A_2^j = \frac{\mu_0 \gamma_S \gamma_{^{13}\mathrm{C}} Z_{\mathrm{eff}}^3}{120 \pi a_0^3} 
  \, N_{nm}^2 \lambda_j,
\end{equation}
where
\begin{align}
   \lambda_x &= 1-\frac{1}{2} \> \beta_{nm}^2,\\
   \lambda_y &= -\frac{1}{2} - \frac{1}{2} \beta_{nm}^2 \sin^2 (3\theta'_{nm}) + 
   \beta_{nm}^2 \cos^2 (3\theta'_{nm}),\\
   \lambda_z &= -\frac{1}{2} + \beta_{nm}^2 \sin^2 (3\theta'_{nm}) - 
   \frac{1}{2} \> \beta_{nm}^2 \cos^2 (3\theta'_{nm}).
\end{align}
We see that the coupling induced by the anisotropic hyperfine interaction is different
in all three spatial directions. Recall that we have labeled our axes such that
$z$, $y$, and $x$ refer to the translational, circumferential, and radial directions,
respectively.
We show typical values for $A_2^j$ in Table \ref{table:hyperfine}.
The direction of strongest hyperfine interaction is radial to the nanotube.
\footnote{The fact that $A_2^x \simeq -2 A_2^y \simeq -2 A_2^z$ for the CNTs in Table
\ref{table:hyperfine} comes directly from the angular integration in the expressions
$\langle \Psi_\sigma | h_2^k | \Psi_{\sigma'}\rangle$, leading to Eq. \eqref{H2}:
the admixture of $\psi_{2p_t}$ and $\psi_{2p_c}$ to the $\pi$-orbital in Eq. \eqref{pi-orbital}
is very small, such that the integration approximately reduces to the $\psi_{2p_\perp}$ term
(the $\psi_{2s}$ term does not contribute by symmetry).}
It is interesting to note the competing signs: the Fermi contact interaction (expressed
via $A_1$) enhances the anisotropic hyperfine interaction along the radial direction,
but reduces it along the circumferential and transverse directions. Furthermore,
the signs of the hyperfine coupling along different directions indicate ferro- and 
antiferromagnetic alignment of the nuclear spins with respect to the electron spin
(in the ground state). We will come back to this in Sec. \ref{sec:knightshift}.

\begin{table}[t]
  \centering
  \begin{tabular}{|l||c|c||c|c||c|c|c|}
    \hline
    & $(10,0)$ & $(20,0)$ & $(5,10)$ & $(9,15)$ & $(5,5)$ & $(10,10)$ & $(20,20)$\\ \hline \hline
    $A_1$ & $0.19$ & $0.05$ & $0.12$ & $0.05$ & $0.26$ & $0.07$ & $0.02$\\ [0.5ex]
    $A_2^x$ & $0.59$ & $0.60$ & $0.59$ & $0.60$ & $0.58$ & $0.60$ & $0.61$\\ [0.5ex]
    $A_2^y$ & $-0.29$ & $-0.30$ & $-0.30$ & $-0.30$ & $-0.29$ & $-0.30$ & $-0.30$\\ [0.5ex]
    $A_2^z$ & $-0.30$ & $-0.30$ & $-0.30$ & $-0.30$ & $-0.30$ & $-0.30$ & $-0.30$\\ [0.5ex]
    \hline
  \end{tabular}
  \caption{\label{table:hyperfine} Estimated hyperfine coupling strengths (in $\mu e \mathrm{V}$)
	for nanotubes of different chiralities $(n,m)$.}
\end{table}

Finally, we address the coupling of electron orbital angular momentum.
Calculating matrix elements of $h_3^k$, it is straightforward to see that
this interaction vanishes identically by applying the operators $L_j$ to the
hydrogenic orbitals appearing in Eq. \eqref{pi-orbital}, then evaluating
$\langle \Psi_{\sigma} | h_3^k | \Psi_{\sigma'} \rangle$ and summing up all
contributions.

\subsection{Graphene}\label{sec:int_graphene}

We consider the `graphene limit' corresponding to let $n,m \rightarrow \infty$
in all expressions in Sec. \ref{sec:int_nanotube}.
Then $\beta_{nm} \rightarrow 0$ and $N_{nm} \rightarrow 1$ and, denoting quantities
related to graphene with a tilde,
\begin{align}
  \label{eq:coupling_graphene}
  \tilde{A}_1 = 0, \quad
  \tilde{A}_2^z = \tilde{A}_2^y = -\frac{\tilde{A}_2^x}{2} = 
  -\frac{\mu_0 \gamma_S \gamma_{^{13}\mathrm{C}} Z_{\mathrm{eff}}^3}{240 \pi a_0^3}.
\end{align}
Inserting numbers, this yields $\tilde{A}_2^z = \tilde{A}_2^y \simeq -0.3 \, \mu e\mathrm{V}$ 
and $\tilde{A}_2^x \simeq 0.6 \, \mu e\mathrm{V}$.
Note that in our notation the $x$-direction is perpendicular to the graphene plane.

Surprisingly, these numbers are not much different from those estimated for $A_2^j$ in 
Sec. \ref{sec:int_nanotube}. Naively, one might have expected a weaker hyperfine interaction
in graphene (as compared to CNTs) due to its flatness and the vanishing contact 
interaction \eqref{ham:contact}. As it turns out, however, even for small CNTs
the contact interaction is only a small correction to the anisotropic hyperfine interaction
\eqref{ham:anisotropic}, so that the latter is the main hyperfine contribution for both
CNTs and flat graphene.

\section{Hyperfine-induced Anisotropic Knight Shift}\label{sec:knightshift}

\begin{figure}[t]
  \centering
  \includegraphics[width=0.7\columnwidth]{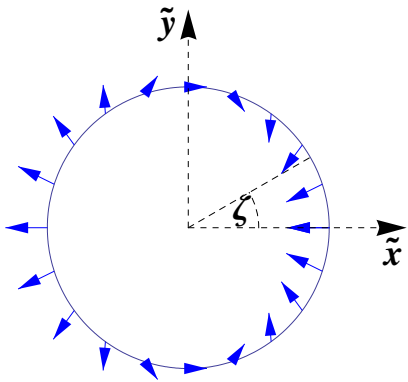}
  \caption{Alignment of the nuclear spins (in the ground state)
  due to the anisotropic Knight shift.
  The electron is assumed to be prepared in the eigenstate of $S^{\tilde{x}}$ with eigenvalue
  $+1/2$, see Eq. \eqref{eq:hfiordering2}, and points along $\tilde{x}$ at each 
  nuclear-spin site.}
  \label{fig:knightshift}
\end{figure}

In Secs. \ref{sec:bonds} and \ref{sec:interactions}, we have formulated the hyperfine
problem in terms of a local coordinate system at each nucleus. In this section,
we want to look at the Knight shift of the $^{13}\mathrm{C}$ nuclear spins due
to hyperfine interaction with the conduction electron. The isotropic Knight shift
due to interaction with both the $sp$-hybridized conduction-band electron and the
$1s$ core electron has been studied in Ref. \onlinecite{Yazyev2005}.
Electron-spin resonance spectra of $^{13}\mathrm{C}$ in $\pi$-electron radicals
have been analyzed in Ref. \onlinecite{Karplus1961}.

From our considerations in Sec. \ref{sec:int_nanotube} it is clear that the Knight
shift induced by the anisotropic hyperfine interaction ($\sim A_2^j/N$)
exceeds the isotropic Knight shift ($\sim A_1/N$) by roughly one order of magnitude
($N$ is the number of nuclei in the dot).
The Knight shift in CNTs (and graphene) is therefore highly anisotropic.

We introduce the following global coordinate
system: $\tilde{x} = x \cos \zeta - y \sin \zeta$, $\tilde{y} = x \sin \zeta + y \cos 
\zeta$, $\tilde{z}=z$, such that the $\tilde{x}\tilde{y}$-plane cuts out a cross section
of the CNT and $\zeta$ is the coordinate describing the position on this cross section
(see Fig. \ref{fig:knightshift}).

The electron is assumed to be prepared in a fixed state and to be delocalized over the
CNT cross section. At each lattice point occupied by a $^{13}\mathrm{C}$ nucleus, the
nuclear spin will align itself in such a way that the hyperfine energy is minimized
in the ground state of the nuclear spins.
This ground state can only be achieved for temperatures that are small with respect to
the energies associated with the Knight shift:
$k_B T < (A_1+A_2^j)/N \sim 1 \, \mathrm{p}e\mathrm{V} \ldots 1 \, \mathrm{n}e\mathrm{V}$,
where $k_B$ is the Boltzmann constant.
We write the Hamiltonian describing the hyperfine interaction of one nucleus with the
delocalized electron as
\begin{equation}
  \label{eq:hfiordering}
  H'_{\mathrm{hf}} = \mathbf{S}^T \, \mathcal{A} \, \mathbf{I}
  = \tilde{\mathbf{S}}^T \, \tilde{\mathcal{A}} \, \tilde{\mathbf{I}},
\end{equation}
where $\mathbf{S}^T=(S^x,S^y,S^z)$, $\mathcal{A}=\mathrm{diag}(A^x,A^y,A^z)$,
$\mathbf{I}=(I^x,I^y,I^z)^T$, $\tilde{\mathbf{S}}^T=\mathbf{S}^T \, R^\dagger$,
$\tilde{\mathcal{A}}=R \, \mathcal{A} \, R^\dagger$,
and $\tilde{\mathbf{I}}=R \, \mathbf{I}$, with the rotation $R$ given by
\begin{equation}
  R =
  \begin{pmatrix}
    \cos \zeta & -\sin \zeta & 0\\
    \sin \zeta & \cos \zeta & 0\\
    0 & 0 & 1
  \end{pmatrix}.
\end{equation}
The operators carrying a tilde hence describe the interaction in the global coordinate
system $(\tilde{x},\tilde{y},\tilde{z})$, and the coupling tensor is given by
\begin{equation*}
  \tilde{\mathcal{A}} =
  \begin{pmatrix}
    A_x \cos^2 \zeta + A_y \sin^2 \zeta & (A_x-A_y) \sin \zeta \, \cos \zeta & 0\\
    (A_x-A_y) \sin \zeta \, \cos \zeta & A_x \cos^2 \zeta + A_y \sin^2 \zeta & 0\\
    0 & 0 & A_z
  \end{pmatrix}.
\end{equation*}
Here, $A_j = A_1 + A_2^j$ is the sum of isotropic and anisotropic hyperfine couplings
along the (local) $j$-direction ($j=x,y,z$), see Eqs.
\eqref{isotropic-coupling} and \eqref{anisotropic-coupling}.
Recall that $A_x>0$ and $A_y, A_z < 0$ for CNTs (see Sec. \ref{sec:int_nanotube}).
For an electron spin in an eigenstate of $S^{\tilde{x}}$, the interaction reads
\begin{equation}
  \label{eq:hfiordering2}
  H_{\mathrm{hf}}^{\tilde{x}} = (A_{\tilde{x} \tilde{x}} I^{\tilde{x}} + A_{\tilde{x} \tilde{y}} 
  I^{\tilde{y}}) S^{\tilde{x}}
\end{equation}
with $A_{\tilde{x} \tilde{x}} = A_x \cos^2 \zeta + A_y \sin^2 \zeta$ and
$A_{\tilde{x} \tilde{y}} = (A_x-A_y) \sin \zeta \, \cos \zeta$.

In Fig. \ref{fig:knightshift} we show the alignment of the nuclear spins
(assuming the nuclear spins to be in their ground state)
due to the anisotropic Knight shift induced by the hyperfine interaction with a conduction
electron whose spin is prepared in the eigenstate of $S^{\tilde{x}}$ with eigenvalue $+1/2$,
i.e., pointing along the $\tilde{x}$-direction at each nuclear site.
We have assumed that the electron is evenly distributed around the CNT cross section,
which can be seen to be justified from the envelope functions calculated 
in Ref. \onlinecite{Bulaev2008}
for semiconducting CNTs subject to a rectangular confinement potential.
We observe an interesting interplay between ferro- and antiferromagnetic coupling
along the two spatial directions, which is a direct consequence of the CNT geometry and 
the strong anisotropy of the hyperfine interaction. 
In particular, the hyperfine interaction does not vanish when we average over the
CNT circumference. This could lead to other interesting effects.
For instance, it has been shown in Ref. \onlinecite{Braunecker2009} that 
in a Luttinger liquid, a non-vanishing average hyperfine field can lead to a transition 
into a helically ordered phase (along the tube axis)
of the nuclear spins below some critical temperature.

\section{Electron-spin decoherence}\label{sec:dynamics}

\subsection{Carbon nanotubes}\label{sec:dyn_nanotube}

Based on our analysis in Sec. \ref{sec:interactions}, the electron-spin dynamics in a CNT
are described by the following hyperfine Hamiltonian:
\begin{equation}
  \label{hf-hamiltonian}
  H_{\mathrm{hf}} = \mathbf{h} \cdot \mathbf{S}, \quad h_j = \sum_k A_k^j I_k^j,
\end{equation}
where $A_k^j = A_j v_0 |\Phi(\mathbf{r}_k)|^2$ and $A_j = A_1 + A_2^j$, see Eqs.
\eqref{isotropic-coupling} and \eqref{anisotropic-coupling}.

We assume that an external magnetic field $B_z$ has been applied along the
symmetry axis of the CNT (see Fig. \ref{fig:nanotube} (b)), 
and that the induced Zeeman splitting between the spin states is larger than the
hyperfine coupling strength: $b = g \mu_B B_z > A_j$. Assuming $g\simeq2$, this corresponds to
very moderate fields of $B_z \gtrsim 5 \, \mathrm{mT}$.
Within this limit, relaxation-induced decoherence is suppressed
by the small parameter $A_j/b$, and the main source of decoherence is pure
dephasing due to nuclear-field fluctuations along the CNT symmetry axis.
The relevant Hamiltonian is given by the $z$-part of Eq. \eqref{hf-hamiltonian} 
and the Zeeman term:
\begin{equation}
  H = (b+h_z) S^z.
\end{equation}
Assuming that all nuclei carry non-zero spin,
the dynamics of the transverse spin is then given by a Gaussian decay with
timescale given by $\sqrt{N}/A_z$, where $N$ is the total number of nuclear 
spins in the quantum dot. \cite{Coish2004}

One advantage of carbon-based nanostructures is the low natural abundance of the 
spin-carrying isotope $^{13}\mathrm{C}$. In order to investigate this advantage,
we allow for arbitrary $^{13}\mathrm{C}$ abundances: we denote the total number 
of nuclei in the quantum dot (defined via the envelope function, see below) by $N$,
and by $N_{12}$ and $N_{13}$ the number of $^{12}\mathrm{C}$ and $^{13}\mathrm{C}$
nuclei, respectively, such that $N=N_{12}+N_{13}$.
This generalization has two relevant effects:
(i) Summing over all $N$ nuclei, we get $\sum_k A_k^z = \eta \, A_z$, i.e., the total
hyperfine coupling is weakened by the ratio $\eta=N_{13}/N$.
(ii) The polarization $p$ of the nuclear-spin system ($0 \leq p \leq 1$)
is determined only by the distribution of
spin-up and spin-down $^{13}\mathrm{C}$ nuclei, while being unaffected by the 
$^{12}\mathrm{C}$ nuclei. 

\begin{figure}[t]
  \centering
  \includegraphics[width=0.95\columnwidth]{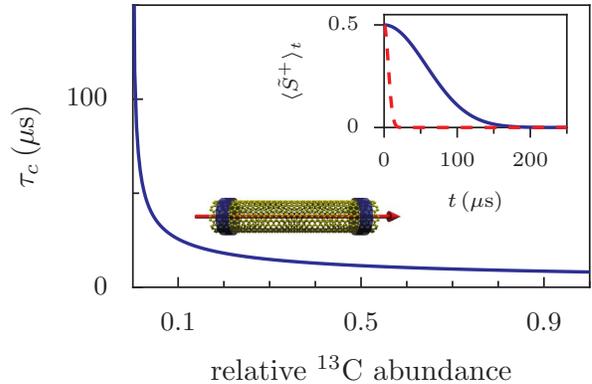}
  \caption{Electron-spin decoherence time $\tau_c$ in a (20,0) CNT as a function of the
  relative $^{13}\mathrm{C}$ abundance $N_{13}/N$, under the condition of a magnetic
  field $B_z \gtrsim 5 \, \mathrm{mT}$ along the symmetry axis of the CNT. We have
  chosen a completely unpolarized nuclear system ($p=0$) and $N=6 \times 10^5$, 
  see text below Eq. \eqref{coherence-time-cnt}.
  Inset: Electron-spin dynamics in CNTs with $^{13}\mathrm{C}$ abundances of
  1\% (solid blue curve) and 99\% (dashed red curve), in the frame rotating
  with $\omega+pN_{13}$, see Eq. \eqref{spin-dynamics}.}
  \label{fig:coherence_nt}
\end{figure}

For $N_{13} \gg 1$ we can use the central limit theorem (compare with
Ref. \onlinecite{Coish2004}) which yields the following Gaussian dynamics 
for the transverse electron spin (written in the frame rotating with
frequency $(\omega+p \eta A_z/2)/\hbar$, where $\omega = b-b_N$ with the nuclear 
Zeeman energy $b_N = g_N \mu_N B_z$):
\begin{equation}
  \label{spin-dynamics}
   \langle S^+ \rangle_t = \langle S^+ \rangle_0 \> e^{-t^2/\tau_c^2}.
\end{equation}
Here, $S^+ = S^x + iS^y$, with the electron-spin operators $S^j=\sigma_j/2$ and
the Pauli operators $\sigma_j$.
The characteristic timescale for the decay is given by \cite{Coish2004}
\begin{equation}
  \label{coherence-time-cnt}
  \tau_c = \frac{2 \hbar}{\sqrt{1-p^2}} \> \frac{N}{\sqrt{N_{13}} A_z}.
\end{equation}
We can see from Eq. \eqref{coherence-time-cnt} that the decoherence time $\tau_c$ has
an interesting non-linear dependence on the abundance of $^{13}\mathrm{C}$ nuclei.

The total number $N$ of nuclei in the quantum dot is determined by the quantum-dot
confinement. Assuming a rectangular confinement along the symmetry axis
\cite{Bulaev2008} and a (20,0) CNT of length $300 \, \mathrm{nm}$, we estimate 
$N \simeq 6 \times 10^5$.
We show the electron-spin decoherence time $\tau_c$ as a function of the relative
$^{13}\mathrm{C}$ abundance in Fig. \ref{fig:coherence_nt} for a completely unpolarized
nuclear bath ($p=0$). We see that decoherence times of several tens of microseconds
can be expected for higher $^{13}\mathrm{C}$ abundances. For the natural
$^{13}\mathrm{C}$ abundance of about 1\%, we estimate $\tau_c \gtrsim 200 \, \mu \mathrm{s}$
from Eq. \eqref{coherence-time-cnt}.
The inset of Fig. \ref{fig:coherence_nt} shows the Gaussian decay of the spin 
coherence (Eq. \ref{spin-dynamics}) for CNTs containing 1\% (solid line) 
and 99\% (dashed line) $^{13}\mathrm{C}$ nuclei.
We note that further increase of the decoherence time $\tau_c$, say by a factor of $x$, 
would require isotopic purification and the reduction of the natural  
$^{13}\mathrm{C}$ abundance by about a factor of $x^2$. 
The decoherence law in Eqs. \eqref{spin-dynamics} and \eqref{coherence-time-cnt}
of course breaks down at the point when only a few nuclear spins are present in the dot.

\subsection{Graphene}\label{sec:dyn_graphene}

\begin{figure}[t]
  \centering
  \includegraphics[width=0.95\columnwidth]{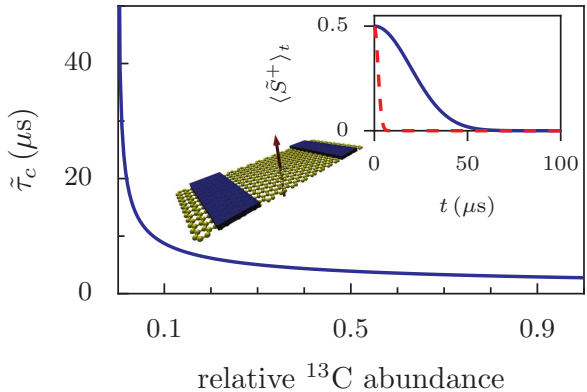}
  \caption{Electron-spin decoherence time $\tau_c$ in graphene as a function of the
  relative $^{13}\mathrm{C}$ abundance $N_{13}/N$, under the condition of a magnetic
  field $B_x \gtrsim 5 \, \mathrm{mT}$ perpendicular to the graphene plane. We have
  chosen $p=0$ and $N=4 \times 10^5$ (see text in Sec. \ref{sec:dyn_graphene}).
  Inset: Electron-spin dynamics in graphene with $^{13}\mathrm{C}$ abundances of
  1\% (solid blue curve) and 99\% (dashed red curve), in the rotating frame.}
  \label{fig:coherence_graphene}
\end{figure}

For an electron confined to a graphene quantum dot, the contact interaction vanishes
identically ($\tilde{A}_1=0$, see Eq. \eqref{eq:coupling_graphene}), and only the
anisotropic hyperfine interaction contributes to spin decoherence.
We assume an externally applied magnetic field $B_x$ perpendicular to the graphene plane
(see Fig. \ref{fig:nanotube} (a)), giving rise to a Zeeman splitting
$\tilde{b} = g_\perp \mu_B B_x$, where $g_\perp \simeq 2$ is the electron g-factor in the
out-of-plane direction. 
If the Zeeman splitting is much larger than the energy
associated with the transverse hyperfine terms, $\tilde{b} \gg \tilde{A}_2^z, \tilde{A}_2^y$,
the electron-spin dynamics are governed by a Gaussian decay similar to
the CNT case \eqref{spin-dynamics}, but with a characteristic timescale given by
\begin{equation}
  \label{coherence-time-graphene}
  \tilde{\tau}_c = \frac{2 \hbar}{\sqrt{1-p^2}} \> \frac{N}{\sqrt{N_{13}} \tilde{A}_2^x}.
\end{equation}
For a quantum dot with width $W=30\, \mathrm{nm}$ and length $L=30\, \mathrm{nm}$,
we estimate the total number of nuclei to be $N \simeq 4 \times 10^5$.
We show $\tilde{\tau}_c$ as a function of the relative $^{13}\mathrm{C}$ abundance
in Fig. \ref{fig:coherence_graphene}: for higher $^{13}\mathrm{C}$ abundances, 
decoherence times $\tilde{\tau}_c \gtrsim 1\mu \mathrm{s}$ can be expected, while for
the natural $^{13}\mathrm{C}$ abundance of 1\%, we estimate $\tilde{\tau_c} \gtrsim
80 \mu \mathrm{s}$.

\subsection{Comparison and discussion}

It is interesting to note that the electron-spin decoherence times in graphene are
shorter than in CNTs. Naively, one would assume that the CNT curvature and the 
associated hybridization would lead to an enhancement of the nuclear-spin interactions
due to the contact interaction \eqref{ham:contact}. As it turns out, however,
the contact interaction in a CNT has a competing sign as compared to the anisotropic
hyperfine interaction along the CNT symmetry axis and, hence, effectively reduces
the total hyperfine coupling strength along the direction of the external magnetic 
field. For ultra-small CNTs it might happen that the total hyperfine coupling along
the symmetry axis approaches zero. 
\footnote{We emphasize that this result is different from Refs. \onlinecite{Yazyev2005,
Semenov2007}, where only the isotropic hyperfine interactions of the nuclear spins with 
the $1s$ and $2s$ electrons have been considered.}
This, however, is beyond the validity of our theory.

For typical CNTs,
the dominant contribution to the nuclear-spin interactions in CNTs comes from the anisotropic
hyperfine interaction. This is because the amount of $s$-orbital admixture in
the hybridized wavefunction of a conduction-band electron is rather small:
on the order of a few percent.

In our considerations throughout this section, we have neglected the hyperfine terms which
are transverse to the externally applied magnetic field, i.e., radial and 
circumferential in the case of a CNT, and in-plane in the case of graphene.
This gives a good first approximation of the decoherence time, as long as the external
magnetic field is large enough to suppress spin flips induced by the hyperfine terms
transverse to the external field. For both CNTs and graphene, magnetic fields of order
$B \gtrsim 5 \, \mathrm{mT}$ are sufficiently strong to achieve this.

\section{Comparison with previous work}\label{sec:comparison}

In Table \ref{table:comparison} we compare our results for the hyperfine interaction in graphene
and CNTs with those given in earlier publications for comparable systems. 
Yazyev's values \cite{Yazyev2008} for the hyperfine interactions in small graphene flakes
were derived from DFT calculations including the $\sigma$-bands. The
values for the anisotropic hyperfine interaction are similar to our results but a 
non-vanishing (and even negative) value for the isotropic hyperfine interaction is reported, 
which is, however, associated with the coupling between the nuclear spins and the
$1s$ core electrons, which is irrelevant for the dephasing of the conduction electron.
Pennington and Stenger \cite{Pennington1996} give estimates for the hyperfine 
interaction in fullerenes and report coupling constants which are slightly larger than
our values. The relatively large value for the isotropic hyperfine interaction can be
explained by the stronger curvature in $\mathrm{C}_{60}$ molecules, leading to stronger
$sp$-hybridization of the electron states, as compared to CNTs.
Goze-Bac \textit{et al.} \cite{GozeBac2002} have estimated the hyperfine interaction in
CNTs based on measurements of the chemical shift of $^{13}\mathrm{C}$.

Churchill \textit{et al.} \cite{Churchill2009a} have estimated the hyperfine
interaction in carbon nanotubes from transport measurements in double quantum dots,
but do not comment on the anisotropy of the interaction.
A hyperfine coupling strength of $\sim 100 \, \mu e \mathrm{V}$ for
pure $^{13}\mathrm{C}$ CNTs is reported in their work, in clear contrast to our results. 
Currently, this discrepancy is not understood. One possible explanation \cite{Trauzettel2009}
for this might be that the theory developed in Ref. \onlinecite{Jouravlev2006}
for standard GaAs quantum dots and used in Ref. \onlinecite{Churchill2009a}
to deduce the hyperfine coupling strength gets modified by the valley degeneracy occurring in 
CNT and thus this could lead to different conclusions.
Another possible explanation \cite{Churchill2009a} could be that the effective electron-nuclear  
interaction gets greatly enhanced by electron-electron interactions and the one-dimensional character 
of the system. A similar renormalization was recently noticed in the context of nuclear magnetic 
ordering in CNT. \cite{Braunecker2009}
Clearly, this is an interesting open problem, which, however, requires separate study.


\begin{table}[t]
  \centering
  \begin{tabular}{|l|c|c|c|}
    \hline
    & $A_1$ [$\mu e \mathrm{V}$] & $A_2^x$ [$\mu e \mathrm{V}$] & 
    $A_2^{y,z}$ [$\mu e \mathrm{V}$]\\ \hline \hline
    our values (CNT) & $0.05$ & $0.6$ & $-0.3$\\ [0.5ex]
    \hline
    our values (Graphene) & $0$ & $0.6$ & $-0.3$\\ [0.5ex]
    \hline \hline
    Yazyev \cite{Yazyev2008} (Graphene flakes) & $-0.2$ & $0.6$ & $-0.3$\\ [0.5ex]
    \hline
    Pennington, Stenger \cite{Pennington1996} ($\mathrm{C}_{60}$) & $0.1$ & $0.9$ & $-0.5$\\ [0.5ex]
    \hline
    Goze-Bac \textit{et al.} \cite{GozeBac2002} (CNT) & $0.04$ & $0.9$ & $-0.5$\\ [0.5ex]
    \hline
  \end{tabular}
  \caption{\label{table:comparison} Comparison of our hyperfine coupling strengths with the values from previous
    publications for comparable systems. 
    We give our values for a (20,0) zigzag nanotube and for graphene.
    See Sec. \ref{sec:comparison} for a detailed discussion.
  }
\end{table}

\section{Conclusions}\label{sec:conclusions}

We have calculated the nuclear-spin interactions and the resulting spin dynamics of an
electron confined to a CNT or graphene quantum dot. In graphene, only the anisotropic
hyperfine interaction couples the electron to the nuclear spins, due to the purely
$p$-type electron wavefunction. In a CNT, curvature induces an $sp$-hybridization of
the electron orbital, opening a new channel of spin decoherence via the Fermi contact
interaction. However, for typical CNTs, the Fermi contact interaction is only
a small correction to the anisotropic hyperfine interaction, the latter being the main source
of nuclear-spin-induced decoherence of the electron spin.
We found the total hyperfine coupling strength of an electron with the $^{13}\mathrm{C}$ 
nuclei to be less than $1 \, \mu e\mathrm{V}$ for both graphene and CNTs quantum dots -- 
about two orders of magnitude smaller than the hyperfine interaction of an electron
in a GaAs or InAs quantum dot.

We have used a simple model for the $sp$-hybridization in CNTs, from which we have
derived the hyperfine interaction.
We have checked, however, that a numerical tight-binding bandstructure 
calculation yields hybridization on the same order of magnitude as the geometrical
approach used in this work. \cite{SchmidtPC} 
Nevertheless, a more rigorous bandstructure calculation
including also the influence of the CNT $\sigma$-bands would be desirable in the future.

For CNTs, we have found an interesting interplay of hyperfine couplings along different spatial
directions, leading to a highly anisotropic Knight shift and an 
alignment of the $^{13}\mathrm{C}$ nuclear spins around the CNT in the ground state of
the nuclear spins. This result is particularly interesting when viewed in context
of the hyperfine-induced nuclear phase transition predicted in Ref. \onlinecite{Braunecker2009} 
for a Luttinger liquid, which requires a non-vanishing mean value of the hyperfine
field. The strong anisotropy of the hyperfine field that we have found in our work
thus gives further evidence that such phase transitions could occur in CNTs.
We emphasize that this anisotropy is present in CNTs of any chirality, in particular also
in metallic CNTs. The results presented here have been used very recently in the
context of transport though CNT quantum dots. \cite{Palyi2009}

Furthermore, we have estimated typical electron-spin decoherence times in CNT
and graphene quantum dots.
We have shown that relaxation-induced decoherence due to nuclear spins
can be suppressed by applying a 
magnetic field of order $5 \, \mathrm{mT}$ to the system, leaving only
pure dephasing due to fluctuations of the nuclear magnetic field. We have estimated
that for a (20,0) zigzag CNT quantum dot containing $N \simeq 6 \times 10^5$ nuclei,
and for a magnetic field applied along the CNT symmetry axis, 
the associated decoherence time is of order $\tau_c \gtrsim 1 \, \mu \mathrm{s}$, 
depending on the relative $^{13}\mathrm{C}$ abundance.
We emphasize that our analytical treatment of the hyperfine problem applies to
CNTs of any chirality.
For a graphene quantum dot containing $N = 4 \times 10^5$ nuclei, and for a magnetic
field applied perpendicular to the graphene plane, the decoherence time is
of order $\tilde{\tau}_c \gtrsim 0.5 \, \mu \mathrm{s}$, again depending on the
relative $^{13}\mathrm{C}$ abundance.

The hyperfine interaction in the systems we have considered here is rather weak.
Therefore, it could, in principle, be that other mechanisms, such as 
spin-orbit interactions,
limit the lifetime of spin-state superpositions on timescales comparable 
to those we have estimated here (see Ref. \onlinecite{Bulaev2008} for details).

The decoherence times we have estimated throughout this work are among the longest
reported so far. This makes quantum dots based on carbon materials attractive
spin-qubit candidates. In particular, the tunability of the average hyperfine 
coupling strength
via the abundance of spin-carrying $^{13}\mathrm{C}$ nuclei could be used to achieve
an optimal balance between a long electron-spin decoherence time and a sufficiently
strong coupling to control the electron-spin state by manipulating the nuclear-spin system.

\begin{acknowledgements}
We acknowledge useful discussions with B. Braunecker, M. J. Schmidt, and
D. Stepanenko, and funding from the Swiss NSF, NCCR Nanoscience, JST ICORP, 
DARPA Quest, QuantumWorks, and the German DFG.
\end{acknowledgements}


\begin{thebibliography}{53}
\expandafter\ifx\csname natexlab\endcsname\relax\def\natexlab#1{#1}\fi
\expandafter\ifx\csname bibnamefont\endcsname\relax
  \def\bibnamefont#1{#1}\fi
\expandafter\ifx\csname bibfnamefont\endcsname\relax
  \def\bibfnamefont#1{#1}\fi
\expandafter\ifx\csname citenamefont\endcsname\relax
  \def\citenamefont#1{#1}\fi
\expandafter\ifx\csname url\endcsname\relax
  \def\url#1{\texttt{#1}}\fi
\expandafter\ifx\csname urlprefix\endcsname\relax\def\urlprefix{URL }\fi
\providecommand{\bibinfo}[2]{#2}
\providecommand{\eprint}[2][]{\url{#2}}

\bibitem[{\citenamefont{Awschalom et~al.}(2002)\citenamefont{Awschalom, Loss,
  and Samarth}}]{AwschalomBook}
\bibinfo{author}{\bibfnamefont{D.~D.} \bibnamefont{Awschalom}},
  \bibinfo{author}{\bibfnamefont{D.}~\bibnamefont{Loss}}, \bibnamefont{and}
  \bibinfo{author}{\bibfnamefont{N.}~\bibnamefont{Samarth}},
  \emph{\bibinfo{title}{Semiconductor Spintronics and Quantum Computing}}
  (\bibinfo{publisher}{Springer-Verlag}, \bibinfo{address}{Berlin},
  \bibinfo{year}{2002}).

\bibitem[{\citenamefont{\ifmmode \check{Z}\else
  \v{Z}\fi{}uti\ifmmode~\acute{c}\else \'{c}\fi{}
  et~al.}(2004)\citenamefont{\ifmmode \check{Z}\else
  \v{Z}\fi{}uti\ifmmode~\acute{c}\else \'{c}\fi{}, Fabian, and
  Das~Sarma}}]{Zutic2004}
\bibinfo{author}{\bibfnamefont{I.}~\bibnamefont{\ifmmode \check{Z}\else
  \v{Z}\fi{}uti\ifmmode~\acute{c}\else \'{c}\fi{}}},
  \bibinfo{author}{\bibfnamefont{J.}~\bibnamefont{Fabian}}, \bibnamefont{and}
  \bibinfo{author}{\bibfnamefont{S.}~\bibnamefont{Das~Sarma}},
  \bibinfo{journal}{Rev. Mod. Phys.} \textbf{\bibinfo{volume}{76}},
  \bibinfo{pages}{323} (\bibinfo{year}{2004}).

\bibitem[{\citenamefont{Awschalom and Flatte}(2007)}]{Awschalom2007}
\bibinfo{author}{\bibfnamefont{D.~D.} \bibnamefont{Awschalom}}
  \bibnamefont{and} \bibinfo{author}{\bibfnamefont{M.~E.}
  \bibnamefont{Flatte}}, \bibinfo{journal}{Nature Physics}
  \textbf{\bibinfo{volume}{3}}, \bibinfo{pages}{153} (\bibinfo{year}{2007}).

\bibitem[{\citenamefont{Loss and DiVincenzo}(1998)}]{Loss1998}
\bibinfo{author}{\bibfnamefont{D.}~\bibnamefont{Loss}} \bibnamefont{and}
  \bibinfo{author}{\bibfnamefont{D.~P.} \bibnamefont{DiVincenzo}},
  \bibinfo{journal}{Phys. Rev. A} \textbf{\bibinfo{volume}{57}},
  \bibinfo{pages}{120} (\bibinfo{year}{1998}).

\bibitem[{\citenamefont{Cerletti et~al.}(2005)\citenamefont{Cerletti, Coish,
  Gywat, and Loss}}]{Cerletti2005}
\bibinfo{author}{\bibfnamefont{V.}~\bibnamefont{Cerletti}},
  \bibinfo{author}{\bibfnamefont{W.~A.} \bibnamefont{Coish}},
  \bibinfo{author}{\bibfnamefont{O.}~\bibnamefont{Gywat}}, \bibnamefont{and}
  \bibinfo{author}{\bibfnamefont{D.}~\bibnamefont{Loss}},
  \bibinfo{journal}{Nanotechnology} \textbf{\bibinfo{volume}{16}},
  \bibinfo{pages}{R27} (\bibinfo{year}{2005}).

\bibitem[{\citenamefont{Hanson et~al.}(2007)\citenamefont{Hanson, Kouwenhoven,
  Petta, Tarucha, and Vandersypen}}]{Hanson2007}
\bibinfo{author}{\bibfnamefont{R.}~\bibnamefont{Hanson}},
  \bibinfo{author}{\bibfnamefont{L.~P.} \bibnamefont{Kouwenhoven}},
  \bibinfo{author}{\bibfnamefont{J.~R.} \bibnamefont{Petta}},
  \bibinfo{author}{\bibfnamefont{S.}~\bibnamefont{Tarucha}}, \bibnamefont{and}
  \bibinfo{author}{\bibfnamefont{L.~M.~K.} \bibnamefont{Vandersypen}},
  \bibinfo{journal}{Rev. Mod. Phys.} \textbf{\bibinfo{volume}{79}},
  \bibinfo{pages}{1217} (\bibinfo{year}{2007}).

\bibitem[{\citenamefont{Burkard et~al.}(1999)\citenamefont{Burkard, Loss, and
  DiVincenzo}}]{Burkard1999}
\bibinfo{author}{\bibfnamefont{G.}~\bibnamefont{Burkard}},
  \bibinfo{author}{\bibfnamefont{D.}~\bibnamefont{Loss}}, \bibnamefont{and}
  \bibinfo{author}{\bibfnamefont{D.~P.} \bibnamefont{DiVincenzo}},
  \bibinfo{journal}{Phys. Rev. B} \textbf{\bibinfo{volume}{59}},
  \bibinfo{pages}{2070} (\bibinfo{year}{1999}).

\bibitem[{\citenamefont{Merkulov et~al.}(2002)\citenamefont{Merkulov, Efros,
  and Rosen}}]{Merkulov2002}
\bibinfo{author}{\bibfnamefont{I.~A.} \bibnamefont{Merkulov}},
  \bibinfo{author}{\bibfnamefont{A.~L.} \bibnamefont{Efros}}, \bibnamefont{and}
  \bibinfo{author}{\bibfnamefont{M.}~\bibnamefont{Rosen}},
  \bibinfo{journal}{Phys. Rev. B} \textbf{\bibinfo{volume}{65}},
  \bibinfo{pages}{205309} (\bibinfo{year}{2002}).

\bibitem[{\citenamefont{Khaetskii et~al.}(2002)\citenamefont{Khaetskii, Loss,
  and Glazman}}]{Khaetskii2002}
\bibinfo{author}{\bibfnamefont{A.~V.} \bibnamefont{Khaetskii}},
  \bibinfo{author}{\bibfnamefont{D.}~\bibnamefont{Loss}}, \bibnamefont{and}
  \bibinfo{author}{\bibfnamefont{L.}~\bibnamefont{Glazman}},
  \bibinfo{journal}{Phys. Rev. Lett.} \textbf{\bibinfo{volume}{88}},
  \bibinfo{pages}{186802} (\bibinfo{year}{2002}).

\bibitem[{\citenamefont{Coish and Loss}(2004)}]{Coish2004}
\bibinfo{author}{\bibfnamefont{W.~A.} \bibnamefont{Coish}} \bibnamefont{and}
  \bibinfo{author}{\bibfnamefont{D.}~\bibnamefont{Loss}},
  \bibinfo{journal}{Phys. Rev. B} \textbf{\bibinfo{volume}{70}},
  \bibinfo{eid}{195340} (\bibinfo{year}{2004}).

\bibitem[{\citenamefont{Johnson et~al.}(2005)\citenamefont{Johnson, Petta,
  Taylor, Yacoby, Lukin, Marcus, Hanson, and Gossard}}]{Johnson2005}
\bibinfo{author}{\bibfnamefont{A.~C.} \bibnamefont{Johnson}},
  \bibinfo{author}{\bibfnamefont{J.~R.} \bibnamefont{Petta}},
  \bibinfo{author}{\bibfnamefont{J.~M.} \bibnamefont{Taylor}},
  \bibinfo{author}{\bibfnamefont{A.}~\bibnamefont{Yacoby}},
  \bibinfo{author}{\bibfnamefont{M.~D.} \bibnamefont{Lukin}},
  \bibinfo{author}{\bibfnamefont{C.~M.} \bibnamefont{Marcus}},
  \bibinfo{author}{\bibfnamefont{M.~P.} \bibnamefont{Hanson}},
  \bibnamefont{and} \bibinfo{author}{\bibfnamefont{A.~C.}
  \bibnamefont{Gossard}}, \bibinfo{journal}{Nature}
  \textbf{\bibinfo{volume}{435}}, \bibinfo{pages}{925} (\bibinfo{year}{2005}).

\bibitem[{\citenamefont{Koppens et~al.}(2005)\citenamefont{Koppens, Folk,
  Elzerman, Hanson, van Beveren, Vink, Tranitz, Wegscheider, Kouwenhoven, and
  Vandersypen}}]{Koppens2005}
\bibinfo{author}{\bibfnamefont{F.~H.~L.} \bibnamefont{Koppens}},
  \bibinfo{author}{\bibfnamefont{J.~A.} \bibnamefont{Folk}},
  \bibinfo{author}{\bibfnamefont{J.~M.} \bibnamefont{Elzerman}},
  \bibinfo{author}{\bibfnamefont{R.}~\bibnamefont{Hanson}},
  \bibinfo{author}{\bibfnamefont{L.~H.~W.} \bibnamefont{van Beveren}},
  \bibinfo{author}{\bibfnamefont{I.~T.} \bibnamefont{Vink}},
  \bibinfo{author}{\bibfnamefont{H.~P.} \bibnamefont{Tranitz}},
  \bibinfo{author}{\bibfnamefont{W.}~\bibnamefont{Wegscheider}},
  \bibinfo{author}{\bibfnamefont{L.~P.} \bibnamefont{Kouwenhoven}},
  \bibnamefont{and} \bibinfo{author}{\bibfnamefont{L.~M.~K.}
  \bibnamefont{Vandersypen}}, \bibinfo{journal}{Science}
  \textbf{\bibinfo{volume}{309}}, \bibinfo{pages}{1346} (\bibinfo{year}{2005}).

\bibitem[{\citenamefont{Cywi\'{n}ski et~al.}(2009)\citenamefont{Cywi\'{n}ski,
  Witzel, and {Das Sarma}}}]{Cywinski2009}
\bibinfo{author}{\bibfnamefont{{\L}.}~\bibnamefont{Cywi\'{n}ski}},
  \bibinfo{author}{\bibfnamefont{W.~M.} \bibnamefont{Witzel}},
  \bibnamefont{and} \bibinfo{author}{\bibfnamefont{S.}~\bibnamefont{{Das
  Sarma}}}, \bibinfo{journal}{Phys. Rev. Lett.} \textbf{\bibinfo{volume}{102}},
  \bibinfo{pages}{057601} (\bibinfo{year}{2009}).

\bibitem[{\citenamefont{Petta et~al.}(2005)\citenamefont{Petta, Johnson,
  Taylor, Laird, Yacoby, Lukin, Marcus, Hanson, and Gossard}}]{Petta2005}
\bibinfo{author}{\bibfnamefont{J.~R.} \bibnamefont{Petta}},
  \bibinfo{author}{\bibfnamefont{A.~C.} \bibnamefont{Johnson}},
  \bibinfo{author}{\bibfnamefont{J.~M.} \bibnamefont{Taylor}},
  \bibinfo{author}{\bibfnamefont{E.~A.} \bibnamefont{Laird}},
  \bibinfo{author}{\bibfnamefont{A.}~\bibnamefont{Yacoby}},
  \bibinfo{author}{\bibfnamefont{M.~D.} \bibnamefont{Lukin}},
  \bibinfo{author}{\bibfnamefont{C.~M.} \bibnamefont{Marcus}},
  \bibinfo{author}{\bibfnamefont{M.~P.} \bibnamefont{Hanson}},
  \bibnamefont{and} \bibinfo{author}{\bibfnamefont{A.~C.}
  \bibnamefont{Gossard}}, \bibinfo{journal}{Science}
  \textbf{\bibinfo{volume}{309}}, \bibinfo{pages}{2180} (\bibinfo{year}{2005}).

\bibitem[{\citenamefont{Trauzettel et~al.}(2007)\citenamefont{Trauzettel,
  Bulaev, Loss, and Burkard}}]{Trauzettel2007}
\bibinfo{author}{\bibfnamefont{B.}~\bibnamefont{Trauzettel}},
  \bibinfo{author}{\bibfnamefont{D.~V.} \bibnamefont{Bulaev}},
  \bibinfo{author}{\bibfnamefont{D.}~\bibnamefont{Loss}}, \bibnamefont{and}
  \bibinfo{author}{\bibfnamefont{G.}~\bibnamefont{Burkard}},
  \bibinfo{journal}{Nature Physics} \textbf{\bibinfo{volume}{3}},
  \bibinfo{pages}{192} (\bibinfo{year}{2007}).

\bibitem[{\citenamefont{Bulaev et~al.}(2008)\citenamefont{Bulaev, Trauzettel,
  and Loss}}]{Bulaev2008}
\bibinfo{author}{\bibfnamefont{D.~V.} \bibnamefont{Bulaev}},
  \bibinfo{author}{\bibfnamefont{B.}~\bibnamefont{Trauzettel}},
  \bibnamefont{and} \bibinfo{author}{\bibfnamefont{D.}~\bibnamefont{Loss}},
  \bibinfo{journal}{Phys. Rev. B} \textbf{\bibinfo{volume}{77}},
  \bibinfo{eid}{235301} (\bibinfo{year}{2008}).

\bibitem[{\citenamefont{Schnez et~al.}(2009)\citenamefont{Schnez, Molitor,
  Stampfer, Guettinger, Shorubalko, Ihn, and Ensslin}}]{Schnez2008}
\bibinfo{author}{\bibfnamefont{S.}~\bibnamefont{Schnez}},
  \bibinfo{author}{\bibfnamefont{F.}~\bibnamefont{Molitor}},
  \bibinfo{author}{\bibfnamefont{C.}~\bibnamefont{Stampfer}},
  \bibinfo{author}{\bibfnamefont{J.}~\bibnamefont{Guettinger}},
  \bibinfo{author}{\bibfnamefont{I.}~\bibnamefont{Shorubalko}},
  \bibinfo{author}{\bibfnamefont{T.}~\bibnamefont{Ihn}}, \bibnamefont{and}
  \bibinfo{author}{\bibfnamefont{K.}~\bibnamefont{Ensslin}},
  \bibinfo{journal}{Appl. Phys. Lett.} \textbf{\bibinfo{volume}{94}},
  \bibinfo{pages}{012107} (\bibinfo{year}{2009}).

\bibitem[{\citenamefont{Ponomarenko et~al.}(2008)\citenamefont{Ponomarenko,
  Schedin, Katsnelson, Yang, Hill, Novoselov, and Geim}}]{Ponomarenko2008}
\bibinfo{author}{\bibfnamefont{L.~A.} \bibnamefont{Ponomarenko}},
  \bibinfo{author}{\bibfnamefont{F.}~\bibnamefont{Schedin}},
  \bibinfo{author}{\bibfnamefont{M.~I.} \bibnamefont{Katsnelson}},
  \bibinfo{author}{\bibfnamefont{R.}~\bibnamefont{Yang}},
  \bibinfo{author}{\bibfnamefont{E.~W.} \bibnamefont{Hill}},
  \bibinfo{author}{\bibfnamefont{K.~S.} \bibnamefont{Novoselov}},
  \bibnamefont{and} \bibinfo{author}{\bibfnamefont{A.~K.} \bibnamefont{Geim}},
  \bibinfo{journal}{Science} \textbf{\bibinfo{volume}{320}},
  \bibinfo{pages}{356} (\bibinfo{year}{2008}).

\bibitem[{\citenamefont{Tans et~al.}(1997)\citenamefont{Tans, Devoret, Dai,
  Thess, Smalley, Geerligs, and Dekker}}]{Tans1997}
\bibinfo{author}{\bibfnamefont{S.~J.} \bibnamefont{Tans}},
  \bibinfo{author}{\bibfnamefont{M.~H.} \bibnamefont{Devoret}},
  \bibinfo{author}{\bibfnamefont{H.}~\bibnamefont{Dai}},
  \bibinfo{author}{\bibfnamefont{A.}~\bibnamefont{Thess}},
  \bibinfo{author}{\bibfnamefont{R.~E.} \bibnamefont{Smalley}},
  \bibinfo{author}{\bibfnamefont{L.~J.} \bibnamefont{Geerligs}},
  \bibnamefont{and} \bibinfo{author}{\bibfnamefont{C.}~\bibnamefont{Dekker}},
  \bibinfo{journal}{Nature} \textbf{\bibinfo{volume}{386}},
  \bibinfo{pages}{474} (\bibinfo{year}{1997}).

\bibitem[{\citenamefont{Bockrath et~al.}(1997)\citenamefont{Bockrath, Cobden,
  McEuen, Chopra, Zettl, Thess, and Smalley}}]{Bockrath1997}
\bibinfo{author}{\bibfnamefont{M.}~\bibnamefont{Bockrath}},
  \bibinfo{author}{\bibfnamefont{D.~H.} \bibnamefont{Cobden}},
  \bibinfo{author}{\bibfnamefont{P.~L.} \bibnamefont{McEuen}},
  \bibinfo{author}{\bibfnamefont{N.~G.} \bibnamefont{Chopra}},
  \bibinfo{author}{\bibfnamefont{A.}~\bibnamefont{Zettl}},
  \bibinfo{author}{\bibfnamefont{A.}~\bibnamefont{Thess}}, \bibnamefont{and}
  \bibinfo{author}{\bibfnamefont{R.~E.} \bibnamefont{Smalley}},
  \bibinfo{journal}{Science} \textbf{\bibinfo{volume}{275}},
  \bibinfo{pages}{1922} (\bibinfo{year}{1997}).

\bibitem[{\citenamefont{Kong et~al.}(2000)\citenamefont{Kong, Zhou, Yenilmez,
  and Dai}}]{Kong2000}
\bibinfo{author}{\bibfnamefont{J.}~\bibnamefont{Kong}},
  \bibinfo{author}{\bibfnamefont{C.}~\bibnamefont{Zhou}},
  \bibinfo{author}{\bibfnamefont{E.}~\bibnamefont{Yenilmez}}, \bibnamefont{and}
  \bibinfo{author}{\bibfnamefont{H.}~\bibnamefont{Dai}},
  \bibinfo{journal}{Appl. Phys. Lett.} \textbf{\bibinfo{volume}{77}},
  \bibinfo{pages}{3977} (\bibinfo{year}{2000}).

\bibitem[{\citenamefont{Minot et~al.}(2004)\citenamefont{Minot, Yaish,
  Sazonova, and McEuen}}]{Minot2004}
\bibinfo{author}{\bibfnamefont{E.~D.} \bibnamefont{Minot}},
  \bibinfo{author}{\bibfnamefont{Y.}~\bibnamefont{Yaish}},
  \bibinfo{author}{\bibfnamefont{V.}~\bibnamefont{Sazonova}}, \bibnamefont{and}
  \bibinfo{author}{\bibfnamefont{P.~L.} \bibnamefont{McEuen}},
  \bibinfo{journal}{Nature} \textbf{\bibinfo{volume}{428}},
  \bibinfo{pages}{536} (\bibinfo{year}{2004}).

\bibitem[{\citenamefont{Jarillo-Herrero
  et~al.}(2004)\citenamefont{Jarillo-Herrero, Sapmaz, Dekker, Kouwenhoven, and
  van~der Zant}}]{Jarillo2004}
\bibinfo{author}{\bibfnamefont{P.}~\bibnamefont{Jarillo-Herrero}},
  \bibinfo{author}{\bibfnamefont{S.}~\bibnamefont{Sapmaz}},
  \bibinfo{author}{\bibfnamefont{C.}~\bibnamefont{Dekker}},
  \bibinfo{author}{\bibfnamefont{L.~P.} \bibnamefont{Kouwenhoven}},
  \bibnamefont{and} \bibinfo{author}{\bibfnamefont{H.~S.~J.}
  \bibnamefont{van~der Zant}}, \bibinfo{journal}{Nature}
  \textbf{\bibinfo{volume}{429}}, \bibinfo{pages}{389} (\bibinfo{year}{2004}).

\bibitem[{\citenamefont{Mason et~al.}(2004)\citenamefont{Mason, Biercuk, and
  Marcus}}]{Mason2004}
\bibinfo{author}{\bibfnamefont{N.}~\bibnamefont{Mason}},
  \bibinfo{author}{\bibfnamefont{M.~J.} \bibnamefont{Biercuk}},
  \bibnamefont{and} \bibinfo{author}{\bibfnamefont{C.~M.}
  \bibnamefont{Marcus}}, \bibinfo{journal}{Science}
  \textbf{\bibinfo{volume}{303}}, \bibinfo{pages}{655} (\bibinfo{year}{2004}).

\bibitem[{\citenamefont{Biercuk et~al.}(2005)\citenamefont{Biercuk, Garaj,
  Mason, Chow, and Marcus}}]{Biercuk2005}
\bibinfo{author}{\bibfnamefont{M.~J.} \bibnamefont{Biercuk}},
  \bibinfo{author}{\bibfnamefont{S.}~\bibnamefont{Garaj}},
  \bibinfo{author}{\bibfnamefont{N.}~\bibnamefont{Mason}},
  \bibinfo{author}{\bibfnamefont{J.~M.} \bibnamefont{Chow}}, \bibnamefont{and}
  \bibinfo{author}{\bibfnamefont{C.~M.} \bibnamefont{Marcus}},
  \bibinfo{journal}{Nano Lett.} \textbf{\bibinfo{volume}{5}},
  \bibinfo{pages}{1267} (\bibinfo{year}{2005}).

\bibitem[{\citenamefont{Cao et~al.}(2005)\citenamefont{Cao, Wang, and
  Dai}}]{Cao2005}
\bibinfo{author}{\bibfnamefont{J.}~\bibnamefont{Cao}},
  \bibinfo{author}{\bibfnamefont{Q.}~\bibnamefont{Wang}}, \bibnamefont{and}
  \bibinfo{author}{\bibfnamefont{H.}~\bibnamefont{Dai}},
  \bibinfo{journal}{Nature Materials} \textbf{\bibinfo{volume}{4}},
  \bibinfo{pages}{745} (\bibinfo{year}{2005}).

\bibitem[{\citenamefont{Sapmaz et~al.}(2006)\citenamefont{Sapmaz, Meyer,
  Beliczynski, Jarillo-Herrero, and Kouwenhoven}}]{Sapmaz2006}
\bibinfo{author}{\bibfnamefont{S.}~\bibnamefont{Sapmaz}},
  \bibinfo{author}{\bibfnamefont{C.}~\bibnamefont{Meyer}},
  \bibinfo{author}{\bibfnamefont{P.}~\bibnamefont{Beliczynski}},
  \bibinfo{author}{\bibfnamefont{P.}~\bibnamefont{Jarillo-Herrero}},
  \bibnamefont{and} \bibinfo{author}{\bibfnamefont{L.~P.}
  \bibnamefont{Kouwenhoven}}, \bibinfo{journal}{Nano Lett.}
  \textbf{\bibinfo{volume}{6}}, \bibinfo{pages}{1350} (\bibinfo{year}{2006}).

\bibitem[{\citenamefont{Onac et~al.}(2006)\citenamefont{Onac, Balestro,
  Trauzettel, Lodewijk, and Kouwenhoven}}]{Onac2006}
\bibinfo{author}{\bibfnamefont{E.}~\bibnamefont{Onac}},
  \bibinfo{author}{\bibfnamefont{F.}~\bibnamefont{Balestro}},
  \bibinfo{author}{\bibfnamefont{B.}~\bibnamefont{Trauzettel}},
  \bibinfo{author}{\bibfnamefont{C.~F.~J.} \bibnamefont{Lodewijk}},
  \bibnamefont{and} \bibinfo{author}{\bibfnamefont{L.~P.}
  \bibnamefont{Kouwenhoven}}, \bibinfo{journal}{Phys. Rev. Lett.}
  \textbf{\bibinfo{volume}{96}}, \bibinfo{pages}{026803}
  (\bibinfo{year}{2006}).

\bibitem[{\citenamefont{Gr\"{a}ber et~al.}(2006)\citenamefont{Gr\"{a}ber,
  Coish, Hoffmann, Weiss, Furer, Oberholzer, Loss, and
  Sch\"{o}nenberger}}]{Graeber2006}
\bibinfo{author}{\bibfnamefont{M.~R.} \bibnamefont{Gr\"{a}ber}},
  \bibinfo{author}{\bibfnamefont{W.~A.} \bibnamefont{Coish}},
  \bibinfo{author}{\bibfnamefont{C.}~\bibnamefont{Hoffmann}},
  \bibinfo{author}{\bibfnamefont{M.}~\bibnamefont{Weiss}},
  \bibinfo{author}{\bibfnamefont{J.}~\bibnamefont{Furer}},
  \bibinfo{author}{\bibfnamefont{S.}~\bibnamefont{Oberholzer}},
  \bibinfo{author}{\bibfnamefont{D.}~\bibnamefont{Loss}}, \bibnamefont{and}
  \bibinfo{author}{\bibfnamefont{C.}~\bibnamefont{Sch\"{o}nenberger}},
  \bibinfo{journal}{Phys. Rev. B} \textbf{\bibinfo{volume}{74}},
  \bibinfo{eid}{075427} (\bibinfo{year}{2006}).

\bibitem[{\citenamefont{J{\o}rgensen et~al.}(2006)\citenamefont{J{\o}rgensen,
  Grove-Rasmussen, Hauptmann, and Lindelof}}]{Jorgensen2006}
\bibinfo{author}{\bibfnamefont{H.~I.} \bibnamefont{J{\o}rgensen}},
  \bibinfo{author}{\bibfnamefont{K.}~\bibnamefont{Grove-Rasmussen}},
  \bibinfo{author}{\bibfnamefont{J.~R.} \bibnamefont{Hauptmann}},
  \bibnamefont{and} \bibinfo{author}{\bibfnamefont{P.~E.}
  \bibnamefont{Lindelof}}, \bibinfo{journal}{Appl. Phys. Lett.}
  \textbf{\bibinfo{volume}{89}}, \bibinfo{pages}{232113}
  (\bibinfo{year}{2006}).

\bibitem[{\citenamefont{Meyer et~al.}(2007)\citenamefont{Meyer, Elzerman, and
  Kouwenhoven}}]{Meyer2007}
\bibinfo{author}{\bibfnamefont{C.}~\bibnamefont{Meyer}},
  \bibinfo{author}{\bibfnamefont{J.~M.} \bibnamefont{Elzerman}},
  \bibnamefont{and} \bibinfo{author}{\bibfnamefont{L.~P.}
  \bibnamefont{Kouwenhoven}}, \bibinfo{journal}{Nano Lett.}
  \textbf{\bibinfo{volume}{7}}, \bibinfo{pages}{295} (\bibinfo{year}{2007}).

\bibitem[{\citenamefont{Kuemmeth et~al.}(2008)\citenamefont{Kuemmeth, Ilani,
  Ralph, and McEuen}}]{Kuemmeth2008}
\bibinfo{author}{\bibfnamefont{F.}~\bibnamefont{Kuemmeth}},
  \bibinfo{author}{\bibfnamefont{S.}~\bibnamefont{Ilani}},
  \bibinfo{author}{\bibfnamefont{D.~C.} \bibnamefont{Ralph}}, \bibnamefont{and}
  \bibinfo{author}{\bibfnamefont{P.~L.} \bibnamefont{McEuen}},
  \bibinfo{journal}{Nature} \textbf{\bibinfo{volume}{452}},
  \bibinfo{pages}{448} (\bibinfo{year}{2008}).

\bibitem[{\citenamefont{Steele et~al.}(2009)\citenamefont{Steele, Gotz, and
  Kouwenhoven}}]{Steele2009}
\bibinfo{author}{\bibfnamefont{G.~A.} \bibnamefont{Steele}},
  \bibinfo{author}{\bibfnamefont{G.}~\bibnamefont{Gotz}}, \bibnamefont{and}
  \bibinfo{author}{\bibfnamefont{L.~P.} \bibnamefont{Kouwenhoven}},
  \bibinfo{journal}{Nature Nanotechnology} \textbf{\bibinfo{volume}{4}},
  \bibinfo{pages}{363} (\bibinfo{year}{2009}).

\bibitem[{\citenamefont{Churchill
  et~al.}(2009{\natexlab{a}})\citenamefont{Churchill, Bestwick, Harlow,
  Kuemmeth, Marcos, Stwertka, Watson, and Marcus}}]{Churchill2009a}
\bibinfo{author}{\bibfnamefont{H.~O.~H.} \bibnamefont{Churchill}},
  \bibinfo{author}{\bibfnamefont{A.~J.} \bibnamefont{Bestwick}},
  \bibinfo{author}{\bibfnamefont{J.~W.} \bibnamefont{Harlow}},
  \bibinfo{author}{\bibfnamefont{F.}~\bibnamefont{Kuemmeth}},
  \bibinfo{author}{\bibfnamefont{D.}~\bibnamefont{Marcos}},
  \bibinfo{author}{\bibfnamefont{C.~H.} \bibnamefont{Stwertka}},
  \bibinfo{author}{\bibfnamefont{S.~K.} \bibnamefont{Watson}},
  \bibnamefont{and} \bibinfo{author}{\bibfnamefont{C.~M.}
  \bibnamefont{Marcus}}, \bibinfo{journal}{Nature Physics}
  \textbf{\bibinfo{volume}{5}}, \bibinfo{pages}{321}
  (\bibinfo{year}{2009}{\natexlab{a}}).

\bibitem[{\citenamefont{Churchill
  et~al.}(2009{\natexlab{b}})\citenamefont{Churchill, Kuemmeth, Harlow,
  Bestwick, Rashba, Flensberg, Stwertka, Taychatanapat, Watson, and
  Marcus}}]{Churchill2009b}
\bibinfo{author}{\bibfnamefont{H.~O.~H.} \bibnamefont{Churchill}},
  \bibinfo{author}{\bibfnamefont{F.}~\bibnamefont{Kuemmeth}},
  \bibinfo{author}{\bibfnamefont{J.~W.} \bibnamefont{Harlow}},
  \bibinfo{author}{\bibfnamefont{A.~J.} \bibnamefont{Bestwick}},
  \bibinfo{author}{\bibfnamefont{E.~I.} \bibnamefont{Rashba}},
  \bibinfo{author}{\bibfnamefont{K.}~\bibnamefont{Flensberg}},
  \bibinfo{author}{\bibfnamefont{C.~H.} \bibnamefont{Stwertka}},
  \bibinfo{author}{\bibfnamefont{T.}~\bibnamefont{Taychatanapat}},
  \bibinfo{author}{\bibfnamefont{S.~K.} \bibnamefont{Watson}},
  \bibnamefont{and} \bibinfo{author}{\bibfnamefont{C.~M.}
  \bibnamefont{Marcus}}, \bibinfo{journal}{Phys. Rev. Lett.}
  \textbf{\bibinfo{volume}{102}}, \bibinfo{pages}{166802}
  (\bibinfo{year}{2009}{\natexlab{b}}).

\bibitem[{\citenamefont{Pennington and Stenger}(1996)}]{Pennington1996}
\bibinfo{author}{\bibfnamefont{C.~H.} \bibnamefont{Pennington}}
  \bibnamefont{and} \bibinfo{author}{\bibfnamefont{V.~A.}
  \bibnamefont{Stenger}}, \bibinfo{journal}{Rev. Mod. Phys.}
  \textbf{\bibinfo{volume}{68}}, \bibinfo{pages}{855} (\bibinfo{year}{1996}).

\bibitem[{\citenamefont{Yazyev}(2008)}]{Yazyev2008}
\bibinfo{author}{\bibfnamefont{O.~V.} \bibnamefont{Yazyev}},
  \bibinfo{journal}{Nano Lett.} \textbf{\bibinfo{volume}{8}},
  \bibinfo{pages}{1011} (\bibinfo{year}{2008}).

\bibitem[{\citenamefont{Saito et~al.}(1998)\citenamefont{Saito, Dresselhaus,
  and Dresselhaus}}]{SaitoBook}
\bibinfo{author}{\bibfnamefont{R.}~\bibnamefont{Saito}},
  \bibinfo{author}{\bibfnamefont{G.}~\bibnamefont{Dresselhaus}},
  \bibnamefont{and} \bibinfo{author}{\bibfnamefont{M.~S.}
  \bibnamefont{Dresselhaus}}, \emph{\bibinfo{title}{Physical Properties of
  Carbon Nanotubes}} (\bibinfo{publisher}{Imperial College Press},
  \bibinfo{year}{1998}).

\bibitem[{\citenamefont{Kleiner and Eggert}(2001)}]{Kleiner2001}
\bibinfo{author}{\bibfnamefont{A.}~\bibnamefont{Kleiner}} \bibnamefont{and}
  \bibinfo{author}{\bibfnamefont{S.}~\bibnamefont{Eggert}},
  \bibinfo{journal}{Phys. Rev. B} \textbf{\bibinfo{volume}{64}},
  \bibinfo{pages}{113402(B)} (\bibinfo{year}{2001}).

\bibitem[{\citenamefont{Fischer et~al.}(2008)\citenamefont{Fischer, Coish,
  Bulaev, and Loss}}]{Fischer2008}
\bibinfo{author}{\bibfnamefont{J.}~\bibnamefont{Fischer}},
  \bibinfo{author}{\bibfnamefont{W.~A.} \bibnamefont{Coish}},
  \bibinfo{author}{\bibfnamefont{D.~V.} \bibnamefont{Bulaev}},
  \bibnamefont{and} \bibinfo{author}{\bibfnamefont{D.}~\bibnamefont{Loss}},
  \bibinfo{journal}{Phys. Rev. B} \textbf{\bibinfo{volume}{78}},
  \bibinfo{eid}{155329} (\bibinfo{year}{2008}).

\bibitem[{\citenamefont{Clementi and Raimondi}(1963)}]{Clementi1963}
\bibinfo{author}{\bibfnamefont{E.}~\bibnamefont{Clementi}} \bibnamefont{and}
  \bibinfo{author}{\bibfnamefont{D.~L.} \bibnamefont{Raimondi}},
  \bibinfo{journal}{J. Chem. Phys.} \textbf{\bibinfo{volume}{38}},
  \bibinfo{pages}{2686} (\bibinfo{year}{1963}).

\bibitem[{\citenamefont{Stoneham}(1972)}]{stoneham}
\bibinfo{author}{\bibfnamefont{A.~M.} \bibnamefont{Stoneham}},
  \emph{\bibinfo{title}{Theory of Defects in Solids}}
  (\bibinfo{publisher}{Oxford University Press}, \bibinfo{year}{1972}),
  \bibinfo{note}{chapter 13}.

\bibitem[{\citenamefont{Latil et~al.}(2001)\citenamefont{Latil, Henrard,
  Goze~Bac, Bernier, and Rubio}}]{Latil2001}
\bibinfo{author}{\bibfnamefont{S.}~\bibnamefont{Latil}},
  \bibinfo{author}{\bibfnamefont{L.}~\bibnamefont{Henrard}},
  \bibinfo{author}{\bibfnamefont{C.}~\bibnamefont{Goze~Bac}},
  \bibinfo{author}{\bibfnamefont{P.}~\bibnamefont{Bernier}}, \bibnamefont{and}
  \bibinfo{author}{\bibfnamefont{A.}~\bibnamefont{Rubio}},
  \bibinfo{journal}{Phys. Rev. Lett.} \textbf{\bibinfo{volume}{86}},
  \bibinfo{pages}{3160} (\bibinfo{year}{2001}).

\bibitem[{\citenamefont{P{\'a}lyi and Burkard}(2009)}]{Palyi2009}
\bibinfo{author}{\bibfnamefont{A.}~\bibnamefont{P{\'a}lyi}} \bibnamefont{and}
  \bibinfo{author}{\bibfnamefont{G.}~\bibnamefont{Burkard}},
  \bibinfo{journal}{arXiv:0908.1054v1}  (\bibinfo{year}{2009}).

\bibitem[{\citenamefont{Paget et~al.}(1977)\citenamefont{Paget, Lampel,
  Sapoval, and Safarov}}]{Paget1977}
\bibinfo{author}{\bibfnamefont{D.}~\bibnamefont{Paget}},
  \bibinfo{author}{\bibfnamefont{G.}~\bibnamefont{Lampel}},
  \bibinfo{author}{\bibfnamefont{B.}~\bibnamefont{Sapoval}}, \bibnamefont{and}
  \bibinfo{author}{\bibfnamefont{V.~I.} \bibnamefont{Safarov}},
  \bibinfo{journal}{Phys. Rev. B} \textbf{\bibinfo{volume}{15}},
  \bibinfo{pages}{5780} (\bibinfo{year}{1977}).

\bibitem[{\citenamefont{Semenov et~al.}(2007)\citenamefont{Semenov, Kim, and
  Iafrate}}]{Semenov2007}
\bibinfo{author}{\bibfnamefont{Y.~G.} \bibnamefont{Semenov}},
  \bibinfo{author}{\bibfnamefont{K.~W.} \bibnamefont{Kim}}, \bibnamefont{and}
  \bibinfo{author}{\bibfnamefont{G.~J.} \bibnamefont{Iafrate}},
  \bibinfo{journal}{Phys. Rev. B} \textbf{\bibinfo{volume}{75}},
  \bibinfo{eid}{045429} (\bibinfo{year}{2007}).

\bibitem[{\citenamefont{Yazyev and Helm}(2005)}]{Yazyev2005}
\bibinfo{author}{\bibfnamefont{O.~V.} \bibnamefont{Yazyev}} \bibnamefont{and}
  \bibinfo{author}{\bibfnamefont{L.}~\bibnamefont{Helm}},
  \bibinfo{journal}{Phys. Rev. B} \textbf{\bibinfo{volume}{72}},
  \bibinfo{pages}{245416} (\bibinfo{year}{2005}).

\bibitem[{\citenamefont{Karplus and Fraenkel}(1961)}]{Karplus1961}
\bibinfo{author}{\bibfnamefont{M.}~\bibnamefont{Karplus}} \bibnamefont{and}
  \bibinfo{author}{\bibfnamefont{G.~K.} \bibnamefont{Fraenkel}},
  \bibinfo{journal}{J. Chem. Phys.} \textbf{\bibinfo{volume}{35}},
  \bibinfo{pages}{1312} (\bibinfo{year}{1961}).

\bibitem[{\citenamefont{Braunecker et~al.}(2009)\citenamefont{Braunecker,
  Simon, and Loss}}]{Braunecker2009}
\bibinfo{author}{\bibfnamefont{B.}~\bibnamefont{Braunecker}},
  \bibinfo{author}{\bibfnamefont{P.}~\bibnamefont{Simon}}, \bibnamefont{and}
  \bibinfo{author}{\bibfnamefont{D.}~\bibnamefont{Loss}},
  \bibinfo{journal}{Phys. Rev. Lett.} \textbf{\bibinfo{volume}{102}},
  \bibinfo{eid}{116403} (\bibinfo{year}{2009}).

\bibitem[{\citenamefont{Goze-Bac et~al.}(2002)\citenamefont{Goze-Bac, Latil,
  Lauginie, Jourdain, Conard, Duclaux, Rubio, and Bernier}}]{GozeBac2002}
\bibinfo{author}{\bibfnamefont{C.}~\bibnamefont{Goze-Bac}},
  \bibinfo{author}{\bibfnamefont{S.}~\bibnamefont{Latil}},
  \bibinfo{author}{\bibfnamefont{P.}~\bibnamefont{Lauginie}},
  \bibinfo{author}{\bibfnamefont{V.}~\bibnamefont{Jourdain}},
  \bibinfo{author}{\bibfnamefont{J.}~\bibnamefont{Conard}},
  \bibinfo{author}{\bibfnamefont{L.}~\bibnamefont{Duclaux}},
  \bibinfo{author}{\bibfnamefont{A.}~\bibnamefont{Rubio}}, \bibnamefont{and}
  \bibinfo{author}{\bibfnamefont{P.}~\bibnamefont{Bernier}},
  \bibinfo{journal}{Carbon} \textbf{\bibinfo{volume}{40}},
  \bibinfo{pages}{1825} (\bibinfo{year}{2002}).

\bibitem[{\citenamefont{Trauzettel and Loss}(2009)}]{Trauzettel2009}
\bibinfo{author}{\bibfnamefont{B.}~\bibnamefont{Trauzettel}} \bibnamefont{and}
  \bibinfo{author}{\bibfnamefont{D.}~\bibnamefont{Loss}},
  \bibinfo{journal}{Nature Physics} \textbf{\bibinfo{volume}{5}},
  \bibinfo{pages}{317} (\bibinfo{year}{2009}).

\bibitem[{\citenamefont{Jouravlev and Nazarov}(2006)}]{Jouravlev2006}
\bibinfo{author}{\bibfnamefont{O.~N.} \bibnamefont{Jouravlev}}
  \bibnamefont{and} \bibinfo{author}{\bibfnamefont{Y.~V.}
  \bibnamefont{Nazarov}}, \bibinfo{journal}{Phys. Rev. Lett.}
  \textbf{\bibinfo{volume}{96}}, \bibinfo{eid}{176804} (\bibinfo{year}{2006}).

\bibitem[{\citenamefont{Schmidt}()}]{SchmidtPC}
\bibinfo{author}{\bibfnamefont{M.~J.} \bibnamefont{Schmidt}},
  \bibinfo{note}{private communication}.

\end{thebibliography}
\end{document}